\documentclass[10pt,aps,prl,amsmath,amssymb,twocolumn,showpacs,byrevtex,superscriptaddress]{revtex4}
\usepackage{graphicx}
\begin{document}
\title{Non-Gaussian energy landscape of a simple model for strong network-forming liquids:
Accurate evaluation of the configurational entropy.}


\author{A. J. Moreno}
\affiliation{Dipartimento di Fisica and CNR-INFM-SMC,
Universit\`a di Roma `La Sapienza', P.le. A. Moro 2, I-00185 Roma, Italy}
\affiliation{Donostia International Physics Center, Paseo Manuel de Lardizabal 4,
E-20018 San Sebasti\'{a}n, Spain}
\author{I. Saika-Voivod}
\affiliation{Dipartimento di Fisica,
Universit\`a di Roma `La Sapienza', P.le. A. Moro 2, I-00185 Roma, Italy}
\affiliation{\mbox{Deparment of Chemistry, University of Saskatchewan, 
110 Science Place, Saskatoon, SK, S7N 5C9, Canada}}

\author{E. Zaccarelli}
\affiliation{\hspace{-1mm}Dipartimento di Fisica and CNR-INFM-SOFT,
Universit\`a di Roma `La Sapienza', P.le. A. Moro 2, I-00185 Roma, Italy}    
\affiliation{ISC-CNR, Via dei Taurini 19, I-00185 Roma, Italy}    

\author{E. La Nave}
\affiliation{\hspace{-1mm}Dipartimento di Fisica and CNR-INFM-SOFT,
Universit\`a di Roma `La Sapienza', P.le. A. Moro 2, I-00185 Roma, Italy}
\affiliation{ISC-CNR, Via dei Taurini 19, I-00185 Roma, Italy}    

\author{ S. V. Buldyrev}
\affiliation{Yeshiva University, Department of Physics, 500 W 185th Street New York, NY 10033, USA}

\author{P. Tartaglia}
\affiliation{Dipartimento di Fisica and CNR-INFM-SMC,
Universit\`a di Roma `La Sapienza', P.le. A. Moro 2, I-00185 Roma, Italy}
\author{F. Sciortino}
\affiliation{\hspace{-1mm}Dipartimento di Fisica and CNR-INFM-SOFT,
Universit\`a di Roma `La Sapienza', P.le. A. Moro 2, I-00185 Roma, Italy}
\begin{abstract}

We present a numerical study of the
statistical properties of the potential energy landscape of a simple model for strong network-forming liquids. 
The model is a system of spherical particles interacting through a square well potential,
with an additional constraint that limits the maximum number of bonds, $N_{\rm max}$, per particle. 
Extensive simulations have been carried out as a function of temperature,
packing fraction, and $N_{\rm max}$. The dynamics of this model are characterized by  
Arrhenius temperature dependence of the transport coefficients and by nearly exponential
relaxation of dynamic correlators,  i.e. features defining strong glass-forming liquids. 
This model has two important features: (i)  landscape basins can be associated with bonding patterns;
(ii) the configurational volume of the basin can be evaluated in a formally exact way,
and numerically with arbitrary precision.  
These features allow us to evaluate  the number of different topologies
the bonding pattern can adopt. We find that the number of fully bonded configurations, 
i.e. configurations in which all particles are bonded to $N_{\rm max}$ neighbors, is extensive, 
suggesting that the configurational entropy of the low temperature fluid is finite. 
We also evaluate the energy dependence of the configurational entropy close to the
fully bonded state, and show that it follows a logarithmic functional form, differently from 
the quadratic dependence characterizing fragile liquids. We suggest that 
the presence of a discrete energy scale, provided by the particle bonds, and the intrinsic 
degeneracy of fully bonded disordered networks differentiates strong from fragile behavior. 
\end{abstract}

\maketitle

\begin{center}
\bf{I. INTRODUCTION}
\end{center}

When a liquid is fastly supercooled into a metastable state under the melting point,
its structural relaxation time, $\tau$, increases over 13 orders of magnitude with decreasing temperature, $T$. 
Below some given temperature, equilibration is not possible within laboratory
time scales and the system becomes a glass \cite{debenedettibook,binder}. The  glass transition
temperature, $T_{\rm g}$, is operationally defined as that where $\tau=100$ seconds,
or the viscosity $\eta=10^{13}$ poise.

Angell has introduced a useful classification scheme \cite{fragdef}
for glass-forming liquids. According to the definition of kinetic fragility, a liquid is classified
as `strong' or `fragile' depending on how fast its relaxation time
increases when approaching $T_{\rm g}$. Liquids that show a weak dependence,
well described by an Arrhenius law $\tau \propto \exp(A/T)$, with $A$ a temperature independent quantity,
are classified as strong. Strong liquids form open network structures
that do not undergo strong structural changes with decreasing temperature. 
On the contrary, the dynamics of fragile liquids, as
many polymeric or low molecular weight organic liquids,
where interactions show a less directional character, are more sensitive to temperature changes,
and relaxation times show strong deviations from Arrhenius behavior \cite{bohmer,martinez}.
Several empirical functions have been proposed for the $T$ dependence
of $\tau$ in fragile liquids, the Vogel-Tammann-Fulcher (VTF) law,
$\tau \propto \exp[A/(T-T_{0})]$, having gained more acceptance \cite{vtf}.
In this equation $T_0$ is the VTF temperature.

Kauzmann noted \cite{kauzmann} that, when extrapolating to low temperature the experimental
$T$ dependence of the configurational entropy, $S_{\rm conf}$, 
the latter became zero at a certain temperature $T_{\rm K}$ (`Kauzmann temperature')
somewhere below $T_{\rm g}$.  Experiments \cite{debenedettibook,notetk} often provide the result
$T_{\rm K} \approx T_0$. 
Given the Arrhenius character of strong liquids,
this comparison would also suggest that $T_{\rm K}$ is zero for these systems.

However, it must be stressed that the values of $T_{\rm K}$ and $T_{0}$ are the result 
of an {\it extrapolation} of experimental data, which in principle
is not necessarily correct. In particular, an extrapolation below $T_{\rm K}$ would
lead to an `entropy catastrophe': a disordered liquid state with less entropy than 
the ordered crystal. In practice, the {\it equilibrium} liquid
state at the putative $T_{\rm K}$ is never reached in experiments, because the
liquid falls {\it out of equilibrium} at $T_{\rm g} > T_{\rm K}$.
The fate of the configurational entropy in an ideal situation
where arbitrarely long equilibration time scales could be accessed is one of the key
(and controversial) questions associated to  the glass transition problem.
One solution states that crystallization is unavoidable when approaching $T_{\rm K}$ 
in equilibrium \cite{kauzmann,cavagna}. It has also been proposed  that $S_{\rm conf}$  
changes its functional form below $T_{\rm g}$, remaining always positive \cite{stillingertk}. 
Another solution is that $S_{\rm conf}$ reaches zero at $T_{\rm K}$ and remains constant 
below it \cite{gibbsdimarzio,derrida,wolynes,mezard}. 

Some insight into this latter question, in the physical origin of the fragility,
and in general in the relation between dynamic and thermodynamic properties
of glass-forming liquids, can be obtained by investigating the potential energy landscape (PEL)
\cite{goldstein,stillinger-pel,stillinger-rev,wales,angell95,jstat}, i.e., the topology of the potential 
energy of the liquid $U = U({\bf r}^{N})$. According to the inherent structure (IS) formalism 
introduced by Stillinger and Weber
\cite{stillinger-pel}, the PEL is partitioned into basins of attraction around the local minima of $U$.
These minima are commonly known as the `inherent structures'.
The free energy is obtained as a sum of a `configurational' contribution, resulting from
the distribution and multiplicity of the different IS's, and another `vibrational'
contribution, resulting from the configurational volume available within the basin around each individual IS.
The introduction of the IS formulation has motivated a great theoretical
and computational effort in order to understand the connection between the statistical
properties of the PEL and the dynamic behavior of supercooled liquids
\cite{sastrynat98,st,buchner1,angelani00,grigera,scala,fdt01,sastry01,debenedetti01,voivod01,middleton,lanave01,
starr01,press,mossaotp,keyes,fabricius,angelani03,doliwa,vogel,denny,newheuer,ruocco,angelanientro,attili}, 
and nowadays has become a key methodology in the field of the glass transition.

From a series of numerical investigations in models of fragile liquids, it is well established that
for such systems the distribution of inherent structures is well described by a Gaussian function,
at least in the energy range that can be probed within
the equilibration times permitted by computational resources \cite{st,sastry01,attili}.
It can be formally proved \cite{derrida,heuer00}
that for a Gaussian landscape the energy of the average visited IS depends linearly
on $T^{-1}$, a result that has been verified in several numerical studies of fragile liquids
\cite{sastry01,starr01,press,buchner1,mossaotp}.
Recent studies of the atomistic BKS model for silica \cite{voivod01,newheuer},
the archetype of strong liquid behavior, have shown
instead that deviations from Gaussianity take place
in the low energy range of the landscape, and indeed, at low temperature,
the average inherent structure progressively deviates from linearity in $T^{-1}$.
The existence of a lower energy cut-off and a discrete energy scale,
as it would be expected for a connected network of bonds,
has been proposed as the origin of such deviations from Gaussianity \cite{newheuer}.
These investigations also suggest that $S_{\rm conf}$ cannot
be extrapolated to zero at a finite $T$, and as a consequence, strong liquids
would not show a finite $T_{\rm K}$.  However, long equilibration times prevent
the determination of the  lowest-energy state and its degeneracy, and 
therefore an unambiguous confirmation of this result. 

Recently, we have proposed a minimal model  which we believe
capture the essence of the prototype strong liquid  behavior. In this model \cite{moreno}, 
particles interact via a spherical square-well potential with an additional constraint on the 
maximum number of bonded neighbors, $N_{\rm max}$, per particle. The lowest-energy state is 
the fully bonded network and its energy is thus unambiguously known.   
Within the equilibration times permitted by present day numerical resources, 
configurations with more than 98 per cent of the bonds formed can be properly thermalized. 
As a result,  no extrapolations are required to determine the low $T$ behavior.
This model is particularly indicated for studying the statistical properties of the landscape.
In analogy with the Stillinger-Weber formalism,  we propose to partition the configuration
space in  basins which, for the present case, can be associated with bonding patterns.
A precise definition of the volume in configuration space associated to each bonding pattern 
can be provided, since a basin is characterized by a flat surface with energy 
proportional to the number of bonds.  Crossing between different basins can be associated 
to bond-breaking or bond-forming events. 
In contrast to other systems previously investigated, the vibrational contribution of the PEL
can be expressed in a formally exact way. No approximation for the shape of the basins
is requested. As we discuss in the following, the precision of the evaluation of the basin
volume, only limited by the numerical accuracy of statistical averages, allows us to evaluate, 
with the same precision, the configurational entropy.
 
In this article we report the study of the statistical properties of the $N_{\rm max}$ model for the cases
$N_{\rm max}=3$, 4 and 5, and for a large range of packing fractions $\phi$, extending the
results limited to a fixed $\phi$ and $N_{\rm max}=4$ previously reported in a short communication \cite{moreno}. 
The range of $\phi$ values here investigated are (0.2-0.35) 
for $N_{\rm max}=3$, (0.3-0.5) for $N_{\rm max}=4$, and $\phi=0.35$ for $N_{\rm max}=5$.
The article is organized as follows. In Section II we introduce the model and
provide computational details. In Section III we briefly summarize the IS formalism and apply it
to the present model. Dynamic and energy landscape features 
are shown and discussed in Section IV. Conclusions are given in Section V.

\begin{center}
{\bf II. MODEL AND EVENT-DRIVEN MOLECULAR DYNAMICS}
\end{center}

The model we investigate is similar in spirit to
one previously introduced by Speedy and Debenedetti \cite{speedymaxval}. 
In the present model, particles interact through a spherical square-well potential
with a constraint on the maximum number of bonds each 
particle can form with neighboring ones. 
Namely, the interaction between any two particles $i$ and $j$ that each have
less than $N_{\rm max}$ bonds to other particles, is given by a spherical square-well
potential of width $\Delta$ and depth $-u_{0}$:
\begin{equation}
V_{ij}(r)=
\begin{cases}
	~~\infty ~~~~~~~\hspace{0.8 mm} r<\sigma    \\ 
        -u_0  ~~~~~~~ \sigma<r<\sigma+\Delta  \\
        ~~~0      ~~~~~~~~ r>\sigma+\Delta ,	
\end{cases}
\label{eq:model1}
\end{equation}
\\
with $r$ the distance between $i$ and $j$. When $\sigma < r < \sigma + \Delta$, 
particles $i$ and $j$ form a bond, unless at least one of them
is already bonded to other $N_{\rm max}$ particles. If this is the case
$V_{ij}(r)$ is simply a hard sphere (HS) interaction:
\begin{equation}
V_{ij}(r)=
\begin{cases}
	~~\infty ~~~~~~~~r<\sigma    \\ 
        ~~~0      ~~~~~~~~\hspace{0.8 mm} r>\sigma.	
\end{cases}
\label{eq:model2}
\end{equation}

In the original model introduced by Speedy and Debenedetti 
\cite{speedymaxval} an angular constraint was imposed by avoiding three particles
bonding loops. In an effort to grasp the basic structural ingredients originating
strong liquid behavior, we have not implemented this additional constraint.
The potential given by Eqs. (\ref{eq:model1}, \ref{eq:model2}), despite its apparent simplicity, 
is not pairwise additive, since at any instant, the interaction between two given particles
does not only depends on $r$, but also on the number of particles bonded to them (if $\sigma < r < \sigma + \Delta$).
Hence, to propagate the system it is requested not only the coordinates and velocities, 
but also the list of bonded interactions. 

We also note that the model is not deterministic.   
Consider a configuration in which particle $i$ is surrounded by  
more than $N_{\rm max}$ other particles within a distance $\sigma< r <\sigma+\Delta$. Of course, only 
$N_{\rm max}$ of these neighbors are bonded to $i$, i.e. feel the square-well interaction.
If one of the bonded neighbors moves out of the square-well interaction range 
(i.e. to a distance $r > \sigma + \Delta$), a bond-breaking process occurs.
According to Eq. (\ref{eq:model1}), a new bond can be formed with one of the other non-bonded particles 
whose position is in the range $\sigma< r <\sigma+\Delta$. If  several candidates
are available ---where a candidate is defined as a particle 
whose distance from $i$ is in the range $\sigma< r <\sigma+\Delta$
{\it and} which is engaged in less than $N_{\rm max}$ bonds to other 
distinct particles --- then one of them is randomly selected to form the new bond with the particle $i$.
Of course, in each bond-breaking/formation process the velocities of the two involved particles
are changed to conserve energy and momentum (see also below).
 
Despite these complications, the model given by Eqs. (\ref{eq:model1}, \ref{eq:model2})
can be considered among the simplest ones for simulating clustering and network formation in fluids \cite{duda,huerta}.  
It does not require three-body angular forces or non-spherical interactions. 
The penalty for retaining spherical symmetry is the complete absence of
geometrical constraints between the bonds.

The maximum number of bonds per particle is controlled by tunning $N_{\rm max}$. 
For a system of $N$ particles, the lowest-energy state, corresponding to the fully bonded network,
has an energy $E_{\rm fb}= -N N_{\rm max}u_0/2$. If $N_{\rm bb}$ is the number of broken bonds 
for a given configuration of the network, the energy of that configuration is given by $E= E_{\rm fb}+N_{\rm bb} u_0$.

The model parameters have been set to $\Delta/(\sigma+\Delta)=0.03$, $u_0=1$ and $\sigma=1$.
In the following, entropy, $S$, will be measured in units of $k_{\rm B}$.  
Setting $k_{\rm B}=1$, potential energy, $E$, and temperature, $T$, are measured in units of $u_0$. 
Distances are measured in units of $\sigma$. Diffusion constants and viscosities are respectively measured
in units of $\sigma(u_{0}/m)^{1/2}$ and $(mu_{0})^{1/2}\sigma^{-2}$.
We have simulated a system of $N=10000$ particles of equal mass $m=1$,
implementing periodic boundary conditions in a cubic cell of length $L_{\rm box}$.
We have evaluated the PEL properties for several values of $N_{\rm max}$,
in a wide range of $T$, and packing fraction $\phi=\pi N/6L_{\rm box}^3$.
The system does not exhibit phase separation for the state points here investigated \cite{gel,gellong,gelgenova}.

Dynamic properties are calculated by molecular dynamics simulations.
We use a standard event-driven algorithm \cite{rapaport} for particles interacting via
discontinuous step potentials. The algorithm calculates, from the particle positions and velocities
at a given instant $t_0$, the instants $t_{\rm coll}$ and positions for all possible collisions between distinct pairs, 
and selects the one which occurs at the smallest $t_{\rm coll}$. Then the system is propagated for a time 
$t_{\rm coll}-t_0$ until the collision occurs. Between collisions, particles move along straight 
lines with constant velocities. Collisions can take place at relative distance $r=\sigma$ (in which case velocities
of colliding particles are reversed according to hard-sphere rules) and, only for bonded interactions, at distance
$\sigma+\Delta$ (in which case velocities are changed in such a way to conserve energy and momentum).
Starting configurations are selected from previously generated hard-sphere configurations with
hard-sphere radius $\sigma+\Delta$. In this way, we  start always from a configuration where no bonds are present.
After thermalization at high $T$ using the potential defined 
in Eqs. (\ref{eq:model1}, \ref{eq:model2}), the system is quenched and equilibrated at the requested temperature. 
Equilibration is achieved when energy and pressure show no drift, and particles have diffused, in average,
several diameters. We also confirm that  dynamic correlators and mean squared displacements show no aging,
i.e., no time shift when being evaluated starting from different time origins. Once the system is equilibrated,
a constant energy run is  performed for production of configurations,  from which 
diffusivities and dynamic correlators are computed. Statistical averages are 
performed over typically 50-100 independent samples.
Standard Monte Carlo (MC) simulations are carried out for the calculation of the 
vibrational contribution of the PEL (see Section III), using previously equilibrated configurations.

\begin{center}
\bf{III. INHERENT STRUCTURE FORMALISM}
\end{center}

If $U=U({\bf r}^{N})$ is the potential energy
of a system of $N$ identical particles of mass $m$, the partition function is given by
the product $Z=Z^{\rm ig}Z^{\rm ex}$, where $Z^{\rm ig}$ is the purely kinetic ideal gas contribution,
and $Z^{\rm ex}$ is the excess contribution resulting from the interaction potential.
The ideal gas contribution to the free energy, $F^{\rm ig} = -k_{\rm B}T\ln Z^{\rm ig}$, is given by:
\begin{equation}
\beta F^{\rm ig}/N = \ln(N/V)+3\ln \Lambda -1,
\label{eq:Fidgas}
\end{equation}
where $V$ is the system volume, $\beta = (k_{\rm B}T)^{-1}$, and $\Lambda = h(2\pi mk_{\rm B}T)^{-1/2}$ 
is the de Broglie wavelength, with $h$ the Planck constant. 
The excess contribution to the partition function is:
\begin{equation}
Z^{\rm ex}=\int d{\bf r}^{N} \exp[-\beta U({\bf r}^{N})].
\end{equation}

The IS formalism introduced by Stillinger and Weber \cite{stillinger-pel}, provides a method for
calculating $Z^{\rm ex}$. According to Stillinger and Weber  the configuration space
is partitioned into basins of attraction of the local minima  of the PEL, the so-called inherent structures.
If $\Omega(E_{\rm IS})$ is the degeneracy of a given IS, i.e., the number of local minima
of $U({\bf r}^{N})$ with potential energy $E_{\rm IS}$, the configurational
entropy is defined as $S_{\rm conf}(E_{\rm IS})= k_{\rm B}\ln[\Omega(E_{\rm IS})]$. 
To properly evaluate $Z^{\rm ex}$, besides the information of the number of distinct IS's, 
it is necessary to evaluate the partition function $Z^{\rm ex}_{\rm vib}(T,E_{\rm IS})$ 
constrained in the volume of each of these basins. The quantity $Z^{\rm ex}_{\rm vib}(T,E_{\rm IS})$ provides
a measure of the configurational volume explored at temperature $T$ by the system 
constrained in the basin of depth $E_{\rm IS}$. Averaging over all basins
with the same depth $E_{\rm IS}$ one obtains \cite{stillinger-pel}: 
\begin{eqnarray}
\nonumber  Z^{\rm ex}_{\rm vib}(T,E_{\rm IS}) = \hspace{3 cm} \\ \frac{1}{\Omega(E_{\rm IS})}
\sum_{{\rm basins}(E_{\rm IS})}\int_{\rm basin} d{\bf r}^{N} e^{-\beta(U-E_{\rm IS})},
\end{eqnarray}
where the notation ``basins($E_{\rm IS}$)'' recalls that the sum is performed over 
all the basins whose potential energy minimum is $E_{\rm IS}$, and each integral 
runs over the configurational volume associated
to the corresponding basin. 
It is convenient to express $Z^{\rm ex}_{\rm vib}(T,E_{\rm IS})$ as:
\begin{equation} 
Z^{\rm ex}_{\rm vib}(T,E_{\rm IS}) = e^{-\beta f^{\rm ex}_{\rm vib}(T,E_{\rm IS})},
\end{equation}
to stress that the `free energy', $f^{\rm ex}_{\rm vib}(T,E_{\rm IS})$, 
commonly referred as the vibrational PEL contribution,  
accounts for the vibrational properties of the system constrained 
in a typical basin of depth $E_{\rm IS}$. 

Within the above framework, the excess partition function is given by:
\begin{equation}
Z^{\rm ex} = \sum_{\rm IS}e^{-\beta[E_{\rm IS}-TS_{\rm conf}(E_{\rm IS})+f^{\rm ex}_{\rm vib}(T,E_{\rm IS})]}.
\end{equation}
In the thermodynamic limit the excess free energy,
$F^{\rm ex}=-k_{\rm B}T\ln Z^{\rm ex}$, can be evaluated as:
\begin{equation}
F^{\rm ex}(T) = E(T)-TS_{\rm conf}(E(T))+f^{\rm ex}_{\rm vib}(T,E(T)),
\label{eq:fexlim} 
\end{equation}
where $E(T)$ is the average IS potential energy at temperature $T$.
Therefore, the excess free energy is the sum of three contributions. 
The fist term in Eq. (\ref{eq:fexlim}) accounts for the average value of the visited local minima
of the potential energy. The second term is related to the degeneracy of
the typical minimum. The third term accounts for the configurational volume of the typical visited basin.   

The IS formalism has been applied in the past to several numerical studies of models of liquids.
Indeed, simulations offer a convenient way to evaluate $F^{\rm ex}(T)$, $E(T)$ and $f^{\rm ex}_{\rm vib}(T,E(T))$,
and to derive, by appropriate substractions, the configurational entropy. The only approximation performed
in these studies refers to the vibrational free energy, which is usually calculated under the
harmonic approximation --- by solving the eigenfrequencies of the Hessian matrix evaluated at the IS ---
or, in the best cases, including anharmonic contributions under the strong assumption \cite{anharmonic} that 
these corrections do not depend on the value of $E_{\rm IS}$.

In the case of the $N_{\rm max}$ model, the evaluation of the free energy in the Stillinger-Weber formalism
is straightforward.  The partition function can be formally written as a sum over 
all distinct bonding patterns --- i.e. over all configurations which cannot be transformed by deformation 
to each other without breaking/forming bonds --- in a way that is formally analogous to the IS approach 
once one identifies a bonding pattern with an IS basin.
Differently from the standard IS approach, the present specific step-wise potential does not
require a minimization procedure to locate the local minimum.  
Each bonding pattern can be associated to  a basin characterized by a flat surface 
with energy proportional to the number of bonds.  Crossing between different basins
requires bond-breaking or bond-forming events.  While in the IS formalism the partition function is
associated to local minima, in the  $N_{\rm max}$ model $Z^{\rm ex}$ is evaluated expanding around all
distinct bonding patterns. The flat surface of the basin and the clear-cut basin boundaries 
make it possible to evaluate the vibrational contribution of the PEL in a formally exact way,  
in contrast to other systems previously investigated. No approximation for the shape of the basins
is requested.  In the $N_{\rm max}$ model, the excess vibrational free energy 
$f^{\rm ex}_{\rm vib}$ is purely entropic.  

To calculate $f^{\rm ex}_{\rm vib}$ we make use of the Perturbed Hamiltonian approach 
introduced by Frenkel and Ladd \cite{frenkelladd,frenkel}, which provides an exact analytical formulation 
for the excess free energy of a given system by integration from a reference Einstein crystal. 
In brief, to calculate the free energy of a system defined by a Hamiltonian $H$, one can 
add a harmonic perturbation,  
$H_\lambda({\bf r}^{N};\lambda_{\rm max})= \lambda_{\rm max}\sum_{i=1}^{N}({\bf r}_i - {\bf r}_i^0)^2$, 
around a disordered configuration ${\bf r}^{N0}=({\bf r}^{0}_{1},...,{\bf r}^{0}_{N})$. 
It can be demonstrated that the excess free energies of the perturbed, $F^{\rm ex}(T;\lambda_{\rm max})$,
and unperturbed, $F^{\rm ex}(T;\lambda=0)$, systems are related as \cite{frenkelladd,frenkel,coluzzi}:
\begin{eqnarray}
F^{\rm ex}(T;\lambda=0)= F^{\rm ex}(T;\lambda_{\rm max}) \nonumber\\
-\int_{-\infty}^{\ln[\lambda_{\rm max}]} \lambda 
\langle \sum_{i=1}^{N} ({\bf r}_i - {\bf r}_i^{ 0})^2 \rangle_{\lambda} d \ln[\lambda].
\label{eq:finteg}
\end{eqnarray}
Brackets denote ensemble average for fixed $\lambda$. Due to the presence of the harmonic perturbation, 
particles in the perturbed system  ($H+H_\lambda$) remain constrained around  ${\bf r}^{N0}$. 
As a consequence,  $\langle \sum_{i=1}^{N} ({\bf r}_i -  {\bf r}_i^{ 0})^2 \rangle_{\lambda}$ is finite. 
For a sufficiently large value of $\lambda_{\rm max}$ (so that $H$ is negligible as compared to $H_\lambda$), 
the perturbed system behaves like an Einstein crystal, i.e., like a system of 3N independent 
harmonic oscillators with elastic constant $\lambda_{\rm max}$. 
The excess free energy for the latter system is given by:
\begin{eqnarray}
\beta F^{\rm ex}(\lambda_{\rm max})/N = 
\beta E/N -\frac{3}{2}\ln\left(\frac{\pi k_{\rm B}T}{\lambda_{\rm max}}\right) \nonumber \\
+1-\ln(N/V).
\label{eq:fexharm}
\end{eqnarray}
In this expression $E$ is the energy of the system in ${\bf r}^{ N0}$. 
If the condition $\lambda_{\rm max}\langle \sum_{i=1}^{N} 
({\bf r}_i - {\bf r}_i^{ 0})^2 \rangle_{\lambda_{\rm max}} =3Nk_{\rm B}T/2$ is fulfilled,
the harmonic limit is recovered and will be also valid for $\lambda'_{\rm max}>\lambda_{\rm max}$. 
Hence, $\lambda_{\rm max}$ can be taken as an upper cut-off for the integration in $\lambda$,
since selecting a larger value $\lambda'_{\rm max}$ leads to trivial
cancelation in Eqs. (\ref{eq:finteg}, \ref{eq:fexharm}).

Next we discuss how the Perturbed Hamiltonian approach can be used for evaluating the 
excess vibrational free energy $f^{\rm ex}_{\rm vib}$ of a typical bonding pattern in 
the $N_{\rm max}$ model. We start from an arbitrary equilibrium configuration ${\bf r}^{ N0}$  
at $T$, and with its associated bonding pattern. We then calculate, via MC,   
the quantity $\langle \sum_{i=1}^{N} ({\bf r}_i -  {\bf r}_i^{ 0})^2 \rangle_{\lambda}$,
imposing the constraint that bonds can neither break nor reform.  In the limit $\lambda \rightarrow 0$
the system samples the volume in configuration space associated to the selected bonding pattern.  
Hence, we can identify $f^{\rm ex}_{\rm vib}(T,E(T))$ with $F^{\rm ex}(T;\lambda = 0)$. By retaining 
the constraint on the bonding pattern, we simulate the perturbed system for several 
values of $\lambda$  to properly evaluate the integrand of Eq. (\ref{eq:finteg})
and estimate the value of $f^{\rm ex}_{\rm vib}(T,E(T))$.  To improve statistics, an average over several 
starting configurations ${\bf r}^{ N0}$  is performed. 

Within the above framework, the corresponding expression
for the excess vibrational entropy, $S^{\rm ex}_{\rm vib}$, is:
\begin{eqnarray}
\frac{S^{\rm ex}_{\rm vib}}{Nk_{\rm B}} = 
\frac{3}{2}\ln\left(\frac{\pi k_{\rm B}T}{\lambda_{\rm max}}\right)
-1+\ln\left(\frac{N}{V}\right)\nonumber \\
+\beta \int_{-\infty}^{\ln[\lambda_{\rm max}]} \lambda 
\langle \sum_{i=1}^{N} ({\bf r}_i - {\bf r}_i^{ 0})^2 \rangle_{\lambda} d \ln[\lambda]. 
\label{eq:svibex}
\end{eqnarray}
We stress that Eq. (\ref{eq:svibex}) is an exact relation for $S^{\rm ex}_{\rm vib}$
for the present $N_{\rm max}$ model.
Hence, the precision in the evaluation of this quantity does not depend on any approximation, but only
on the numerical accuracy of the MC calculation and the statistical average.

The total excess entropy $S^{\rm ex}_{\rm tot}(T)$ can also be
calculated with arbitrary precision by thermodynamic integration
from a reference state at $T=T_{\rm ref}$ where $S^{\rm ex}_{\rm tot}$ is already known. 
One possible choice is to select  as reference point the ideal gas 
and integrate along a path which does not cross any
phase boundary or, in the present case, to integrate from very high temperature. Indeed, at sufficiently 
high $T_{\rm ref}$, the model is equivalent to a hard-sphere system, whose free energy is well known. 
An accurate estimate of the hard-sphere excess entropy is provided
by the Carnahan-Starling formula \cite{carnahan}:
\begin{equation}
\frac{S^{\rm ex}_{\rm HS}}{Nk_{\rm B}} = \frac{\phi(3\phi-4)}{(1-\phi)^{2}}.
\label{eq:carnstar}
\end{equation}
The excess entropy at finite $T$ can then be obtained by integrating along a constant volume $V$ path as
\begin{equation}
S^{\rm ex}_{\rm tot}(T) = S^{\rm ex}_{\rm HS} + 
\int_{T_{\rm ref}}^{T}\left(\frac{\partial E}{\partial T}\right)_{V}\frac{dT}{T}.
\label{eq:stotex}
\end{equation}

From the two accurate evaluations of $S^{\rm ex}_{\rm tot}$ [Eq. (\ref{eq:stotex})]
and $S^{\rm ex}_{\rm vib}$ [Eq. (\ref{eq:svibex})] an accurate estimate of the configurational entropy
$S_{\rm conf} $ can be obtained as  $S_{\rm conf} = S^{\rm ex}_{\rm tot}-S^{\rm ex}_{\rm vib}$.
The resulting $S_{\rm conf} $ can be related to $T$ or, parametrically, to $E(T)$ (i.e., to the number of bonds).  
In the rest of the article, all the thermodynamic functions $\Psi$ will 
be expressed as ``quantity per particle'', but for simplicity of notation we will
drop the factor $1/N$. Hence, in the following ``$\Psi$" will be understood as ``$\Psi/N$''.

\begin{center}
{\bf IV. RESULTS AND DISCUSSION}
\end{center}

\begin{center}
{\bf  a) Dynamics}
\end{center}

\begin{figure}
\includegraphics[width=.4\textwidth]{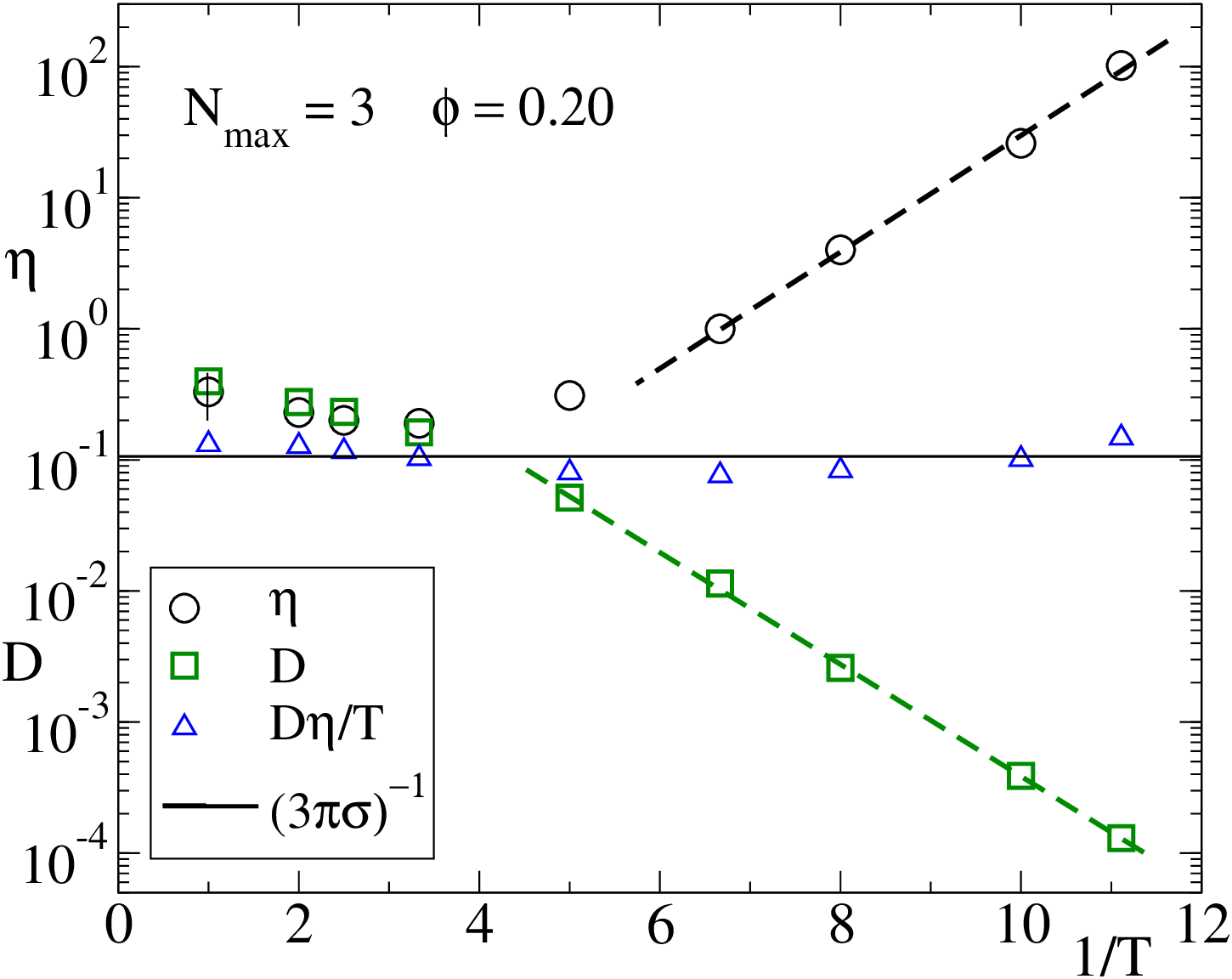}
\newline
\newline
\includegraphics[width=.4\textwidth]{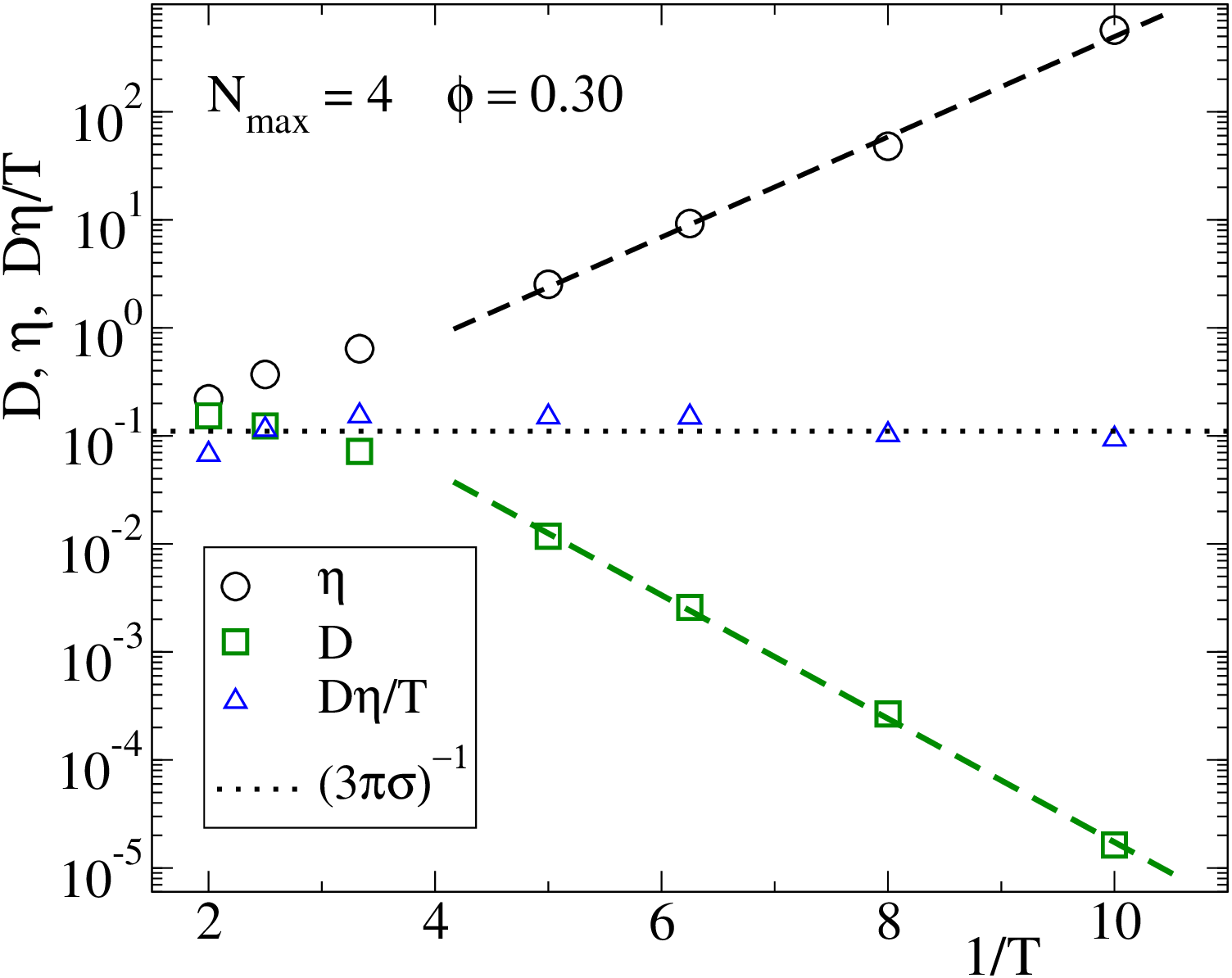}
\newline
\caption{$T$ dependence of the diffusivity $D$, the viscosity $\eta$,
and the product $D\eta/T$ for the cases $N_{\rm max} = 3$, $\phi = 0.20$ (top panel),
and $N_{\rm max} = 4$, $\phi = 0.30$ (bottom panel). Dotted lines correspond
to the expected value $(3\pi\sigma)^{-1}$ from the Stokes-Einstein relation.
Dashed lines are fits to Arrhenius laws. An error bar is included for the viscosity at high $T$.}
\label{fig:diffvisc}
\end{figure}

In this Section we show that  the dynamics in the  $N_{\rm max}$  model   meet the criteria defining 
strong liquids. The key feature is the Arrhenius behavior
of the transport coefficients. Fig. \ref{fig:diffvisc} shows the $T$ dependence 
of the diffusivity $D$ and the 
viscosity $\eta$ for different values of $N_{\rm max}$ and $\phi$. 
The diffusivity is calculated as the
long time limit of $\langle \sum_{i=1}^{N}[{\bf r}_{i}(t)-{\bf r}_{i}(0)]^{2}\rangle/6Nt$.
The viscosity $\eta$ is determined as:
\begin{equation}
\eta = \frac {1}{2V k_{\rm B} T} \lim\limits_{t \rightarrow \infty}
\frac{d}{dt} \langle[A(t)-A(0)]^2\rangle   ,
\label{eq:diffvisc}
\end{equation}
where $A(t)=m \sum_{i=1}^{N} \dot r_{i}^{\alpha} r_{i}^{\beta}$, as explained in \cite{alder}.
Greek symbols denote the $x,y,z$ coordinates of the position, ${\bf r}_{i}$,
of particle $i$. An average is done over all the permutations with $\alpha \neq \beta$.
As shown in Fig. \ref{fig:diffvisc}, at low temperature
both quantities display Arrhenius behavior. The activation energies for $D$ and $\eta$ are 
approximately $u_{0}$, suggesting that all bonds break and reform essentially in an independent way. 
This is consistent with the absence of  angular constrains  in this model.  
Despite the simulations are limited in time by computational resources, 
the fact that the Arrhenius form covers more than
three orders of magnitude and the observed value of the activation energy strongly 
suggest that this functional form will be retained at lower $T$. 
We also find (see Fig. \ref{fig:diffvisc}) that the Stokes-Einstein relation \cite{hansen}, 
$D \eta /T = (3 \pi \sigma)^{-1}$, is fulfilled 
essentially at all temperatures, independently from the $N_{\rm max}$ value.

\begin{figure}
\includegraphics[width=.42\textwidth]{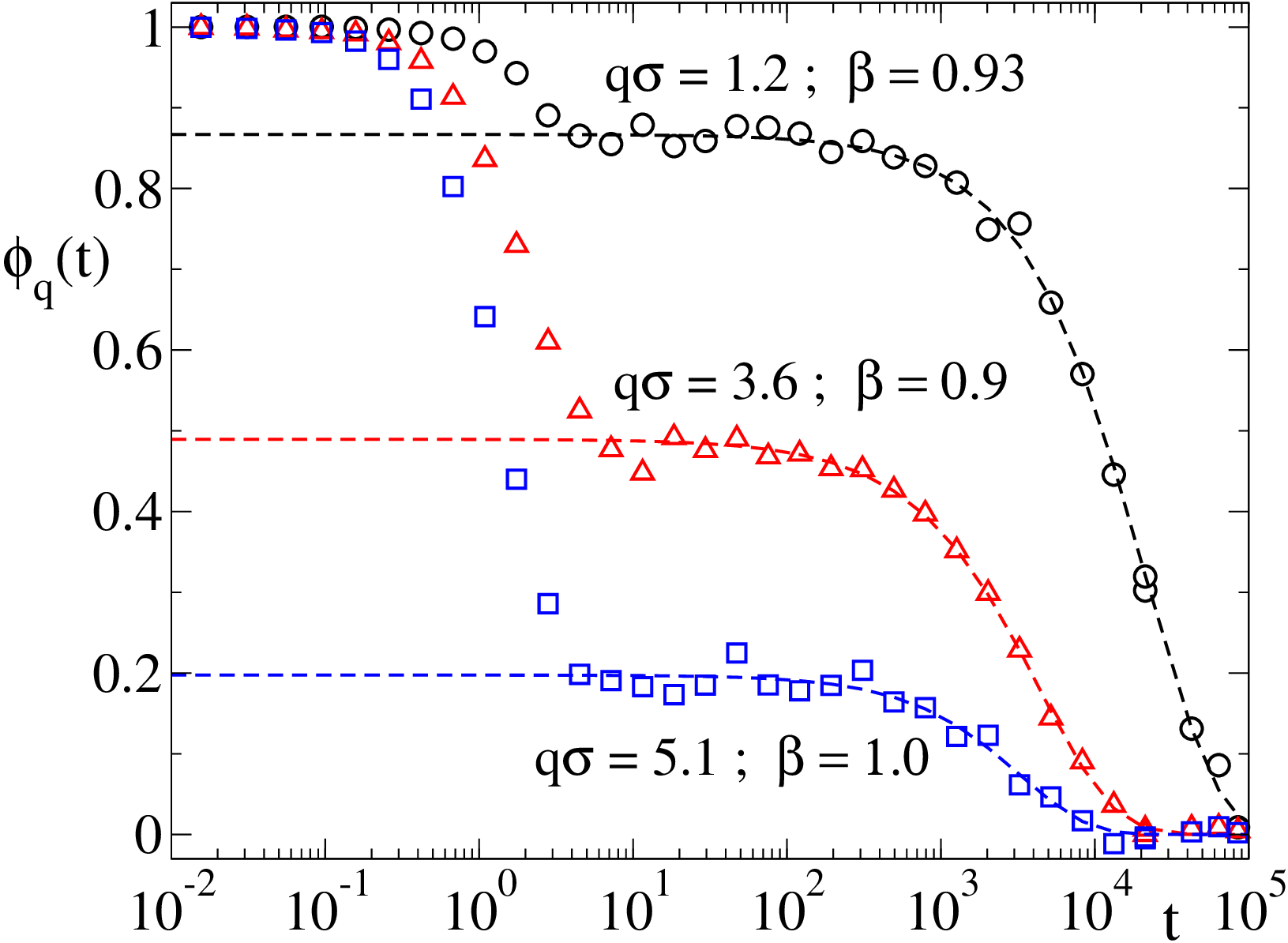}
\newline
\caption{Symbols: Coherent intermediate scattering function for $N_{\rm max}=4$, $\phi=0.30$,
and $T=0.1$, for different values of the wavevector $q$. Dashed lines are KWW fits.
The stretching exponents $\beta$ are indicated for the corresponding $q$'s. }
\label{fig:fsqt}
\end{figure}

Long time decays of dynamic correlators in supercooled states are usually well described
by the phenomenological Kohlrausch-Williams-Watts (KWW) function $\exp[-(t/\tau)^{\beta}]$, where $\tau$
is the corresponding relaxation time and $\beta$ is a stretching exponent which takes values
$0 < \beta < 1$. Experimental evidence for a collection of chemically and structurally
very different glass-forming liquids \cite{bohmer}
shows that the smaller the fragility index (i.e., the closer the system is to
strictly strong behavior), the closer to unity the values of $\beta$ are. 
Figure \ref{fig:fsqt} shows that this is indeed the case for the $N_{\rm max}$ model. 
The long-time dependence of the normalized coherent intermediate scattering function,
$\phi_q(t) = \langle\rho_q(t)\rho_{-q}(0)\rangle/\langle\rho_q(0)\rho_{-q}(0)\rangle$, where
$\rho_q(t)=\sum_{i=1}^N \exp[i {\bf q}\cdot {\bf r_i}(t)]$, can be well described by 
KWW fits with values of $\beta > 0.85$ in all the $q$ range and for all studied $\phi$.
Such a behavior is observed at all $T$ where the system shows Arrhenius behavior.  As shown in
Ref. \cite{bohmer}, these $\beta$ values are very different from the ones  typical of fragile
liquids ($\beta \sim 0.5$).

Results reported  in Figs. \ref{fig:diffvisc} and \ref{fig:fsqt} provide convincing evidence
that Eqs. (\ref{eq:model1}, \ref{eq:model2}) define a simple and satisfactory minimal model
for a strong glass-forming liquid. 

\begin{center}
{\bf b) Energy landscape}
\end{center}

\begin{figure}
\includegraphics[width=.39\textwidth]{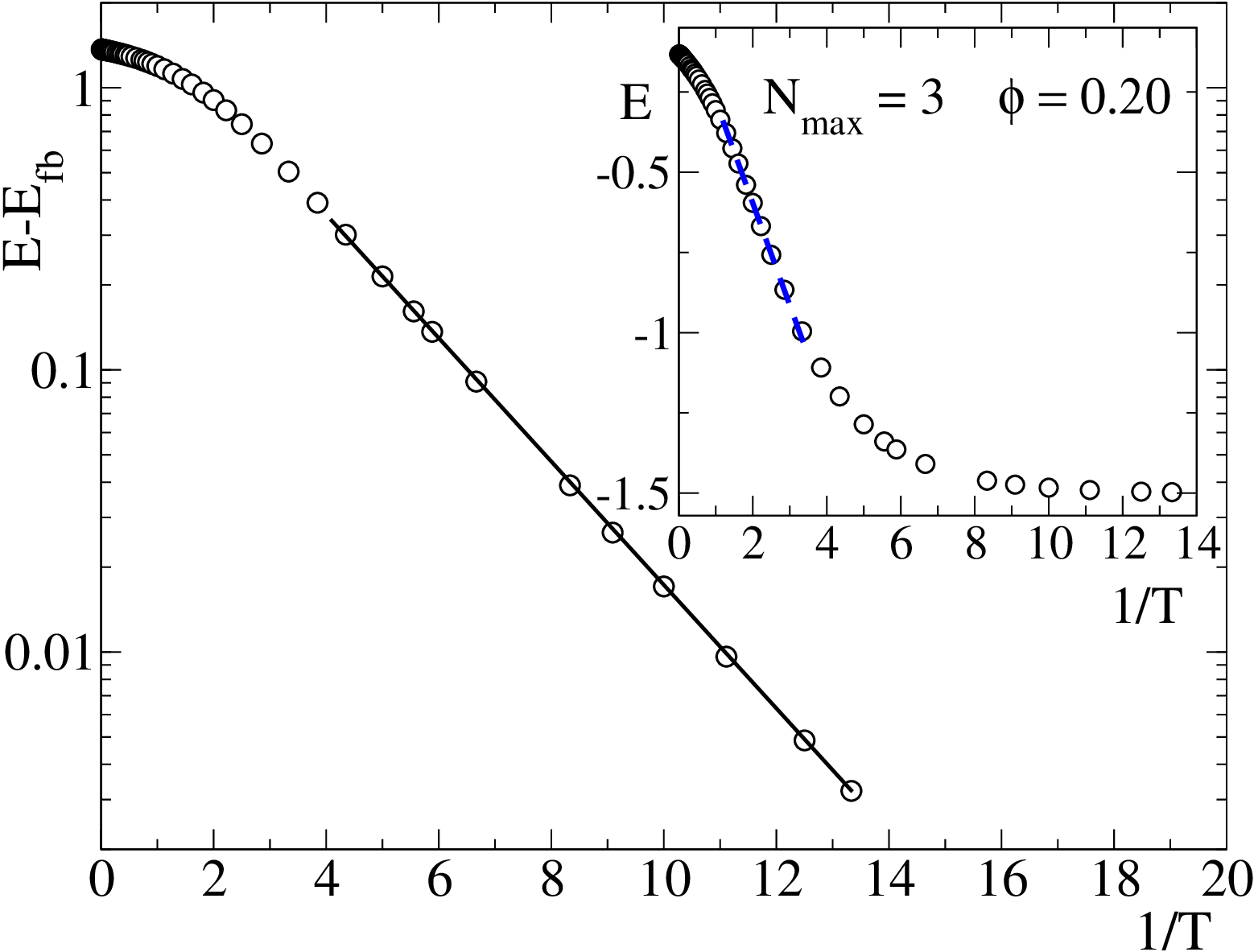}
\newline
\newline
\includegraphics[width=.39\textwidth]{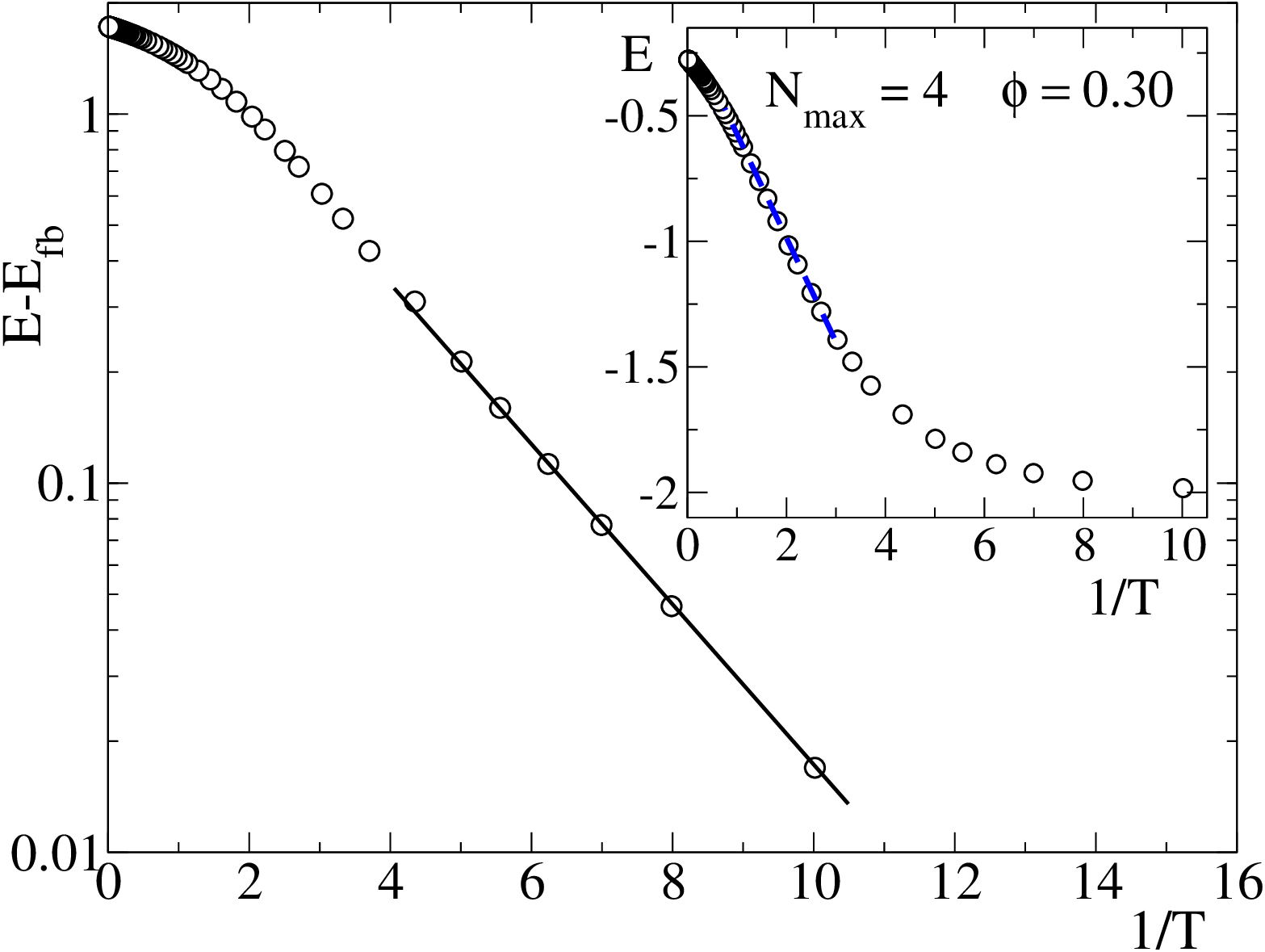}
\newline
\newline
\includegraphics[width=.39\textwidth]{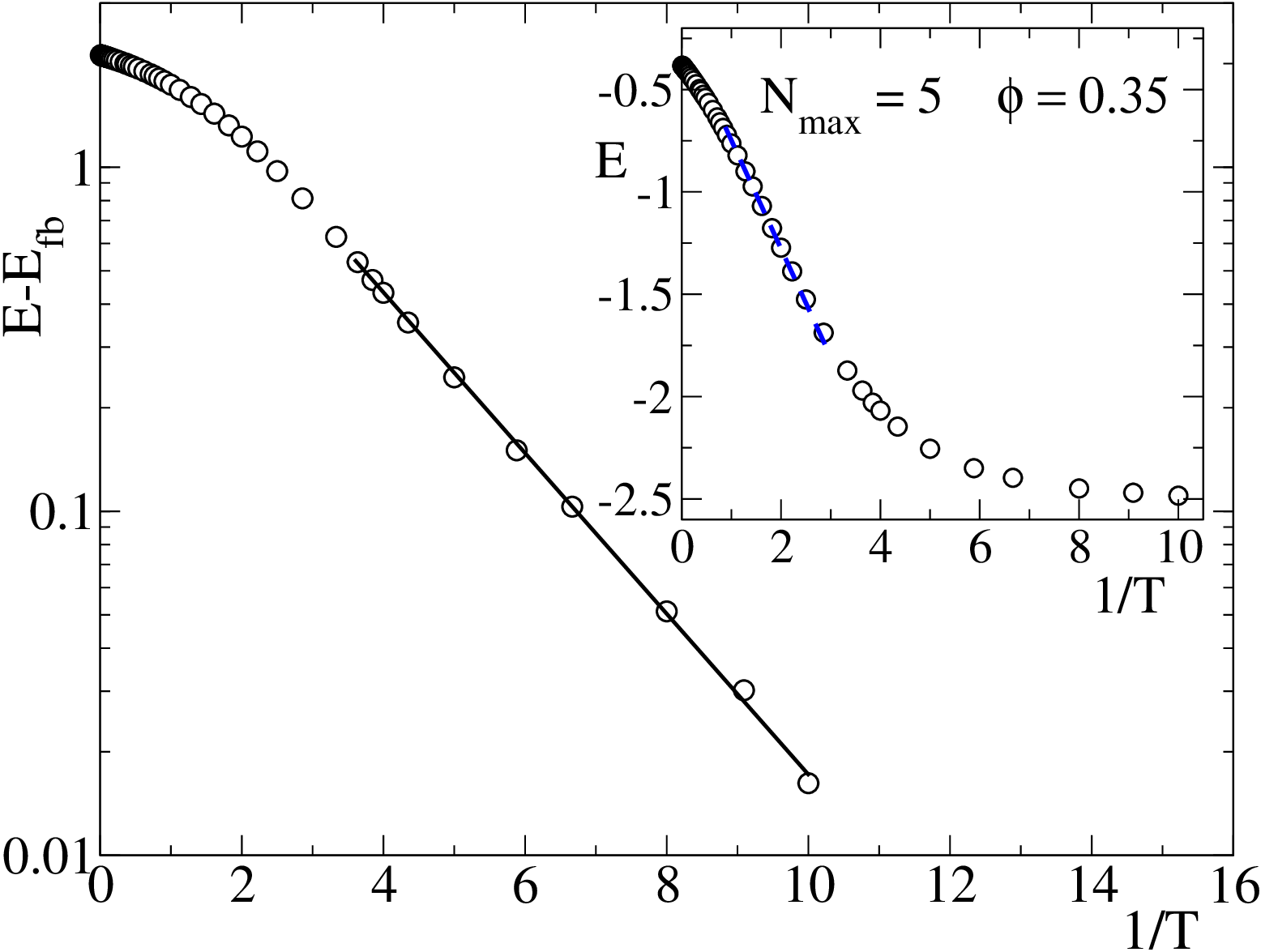}
\newline
\caption{$T$ dependence of the potential energy per particle $E$
for the cases  $N_{\rm max} = 3$, $\phi = 0.20$ (top panel),
$N_{\rm max} = 4$, $\phi = 0.30$ (middle panel),
and $N_{\rm max} = 5$, $\phi = 0.35$ (bottom panel). Full lines at low $T$
are fits to Arrhenius behavior with activation energy $u_0/2$
(see text and Table \ref{tab:ener}).
Dashed lines at high $T$ correspond to linear behavior in $T^{-1}$ (see text).}
\label{fig:ene}
\end{figure}

Fig. \ref{fig:ene} shows the $T$ dependence of the potential energy 
(i.e., the energy of the typical bonding pattern) per particle, $E$. 
In the present model, the potential energy of each configuration coincides with the
energy of the bonding pattern and can be directly associated, in the Stillinger-Weber formalism,
with the IS energy. Within the times accessible by the simulations, 
equilibrium states can be reached
for configurations characterized by a number of broken bonds smaller than  a 2 \%, i.e. the
lowest energy state, $E_{\rm fb}$, is approached from equilibrium simulations.

From the low $T$ behavior of $E$, one sees that the approach to $E_{\rm fb}$ is well described by an Arrhenius law:
\begin{equation}
E-E_{\rm fb} = A_{\infty}^{\rm f} \exp(-\epsilon_{\rm f} /k_{\rm B}T). 
\label{eq:arr}
\end{equation}
The activation energy $\epsilon_{\rm f}$, determined by a free fit, is close to $u_{0}/2$. Indeed,
by forcing the Arrhenius activation energy to be exactly $u_{0}/2$ a satisfactory
representation of the data is recovered, with one simple fitting parameter $A_{\infty}$. 
Figure \ref{fig:ene} shows the result of a fit to $E-E_{\rm fb}  =A_{\infty}\exp(-u_{0}/2k_{\rm B}T)$.
The corresponding best-fitting values, with two ($A^{\rm f}_{\infty}, u_0$) and one ($A_{\infty}$) 
free parameters are reported in Table \ref{tab:ener}.
The observed value $u_{0}/2$ is  consistent with theoretical predictions based on the
thermodynamic perturbation theory developed  by Wertheim \cite{wertheim1} to study association in simple liquids.  
It is not a coincidence since in Wertheim's theory bonds are also geometrically uncorrelated.
Similar values are also predicted by more intuitive recent approaches \cite{sear}.

\begin{table}[tbh]
\begin{center}
\begin{tabular}{lcccccccc}
&$(N_{\rm max},\phi)$ &\ &$A_{\infty}$ &\ &$A_{\infty}^{\rm f}$ &\ &$\epsilon_{\rm f}$    \\
\hline
\\
& (3, 0.20)  &\ & 2.69 &\  & 2.70 &\  & 0.506   \\
\vspace{0.2mm}
& (3, 0.30)  &\ & 1.74 &\  & 1.72 &\  & 0.499   \\
\vspace{0.2mm}
& (3, 0.35)  &\ & 1.38 &\  & 1.37 &\  & 0.498   \\
\vspace{0.2mm}
& (4, 0.30)  &\ & 2.59 &\  & 2.73 &\  & 0.508   \\
\vspace{0.2mm}
& (4, 0.35)  &\ & 2.15 &\  & 2.12 &\  & 0.498   \\
\vspace{0.2mm}
& (4, 0.40)  &\ & 1.70 &\  & 1.67 &\  & 0.498   \\
\vspace{0.2mm}
& (4, 0.45)  &\ & 1.28 &\  & 1.23 &\  & 0.495   \\
\vspace{0.2mm}
& (4, 0.50)  &\ & 0.900 &\  & 0.884 &\  & 0.498   \\
\vspace{0.2mm}
& (5, 0.35)  &\ & 3.14  &\ & 3.73  &\ & 0.539   \\    
\\
\hline
\end{tabular}
\end{center}
\caption{Fit parameters for the low $T$ Arrhenius dependence of
the potential energy $E$ (see text).}
\label{tab:ener}
\end{table}

The clear low $T$ Arrhenius dependence and the explicit value of the activation energy provide
a convenient way to evaluate the $T$ dependence of the energy for lower $T$.  
While in $T$, it might appear as a wide extrapolation procedure, we recall that in $E$ the 
{\it interpolation}  extends only over a small (2\%) range of energies, between the fully bonded
state ($E=E_{\rm fb}$) and the lowest equilibrated state studied in simulations.  Eq. (\ref{eq:arr}) provides
a convenient expression for the low $T$ behavior of $E$ and, by using Eq. (\ref{eq:stotex}),
a way of calculating the total excess entropy down to the fully connected state.

We note on passing that at intermediate temperatures the $T$ dependence of $E$ is consistent
with a $1/T$ law, as expected for a Gaussian distribution of energy levels. The $1/T$ law
crosses to the Arrhenius dependence on cooling. This crossing has been 
also observed in the study of the  $T$ dependence of the IS energy  in a realistic (atomistic)
model for silica \cite{voivod01,newheuer}.

\begin{figure}
\includegraphics[width=.41\textwidth]{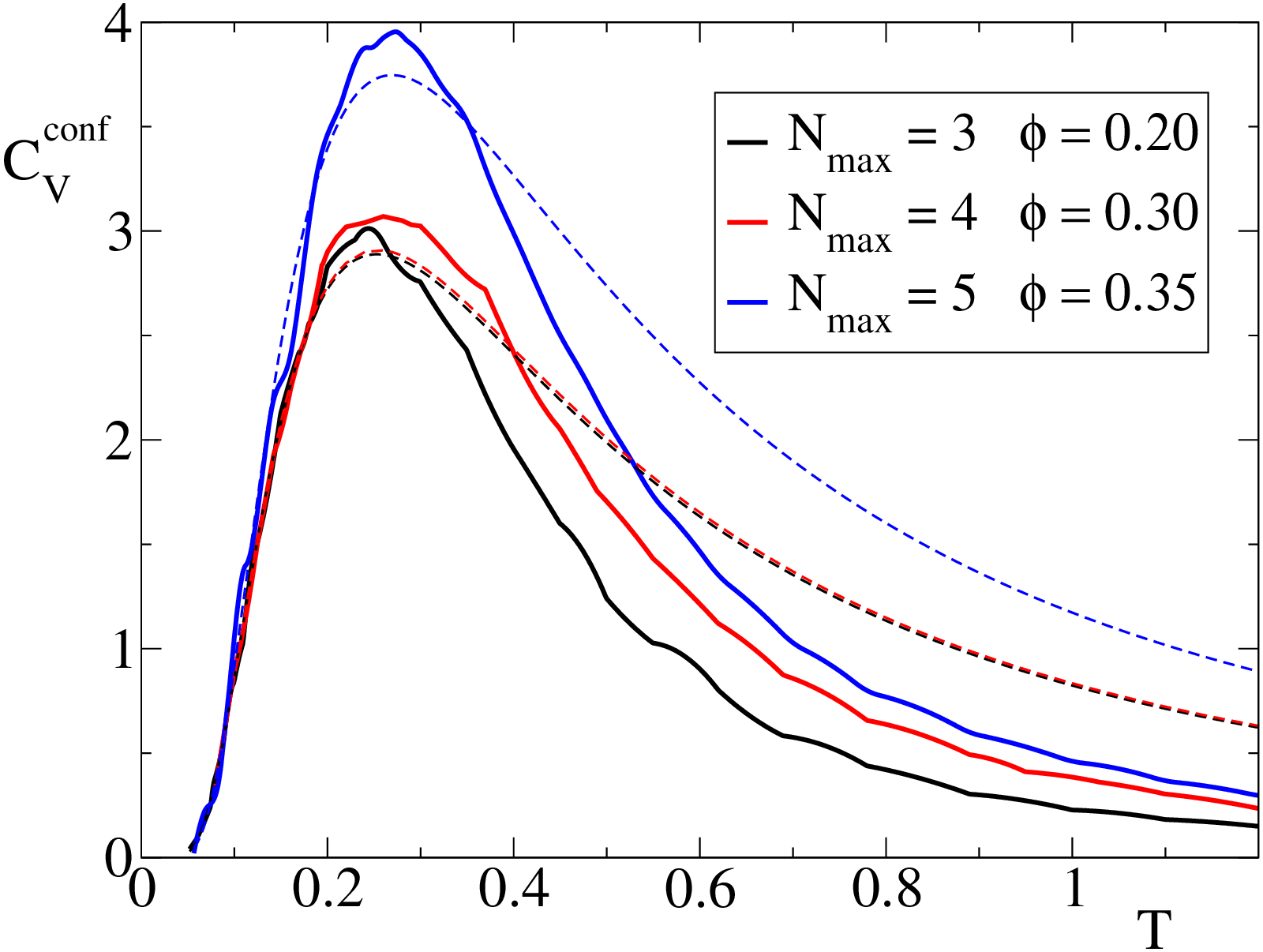}
\newline
\caption{Full lines: $T$ dependence of the isochoric configurational specific heat for several values
of ($N_{\rm max}$, $\phi$). Dashed lines are an extrapolation to high $T$ of the low $T$ behavior
$A^{\rm f}_{\infty}\epsilon_{\rm f}k^{-1}_{\rm B}T^{-2}\exp(-\epsilon_{\rm f}/k_{\rm B}T)$ (see text).}
\label{fig:cvpeak}
\end{figure}

\begin{figure}
\includegraphics[width=.4\textwidth]{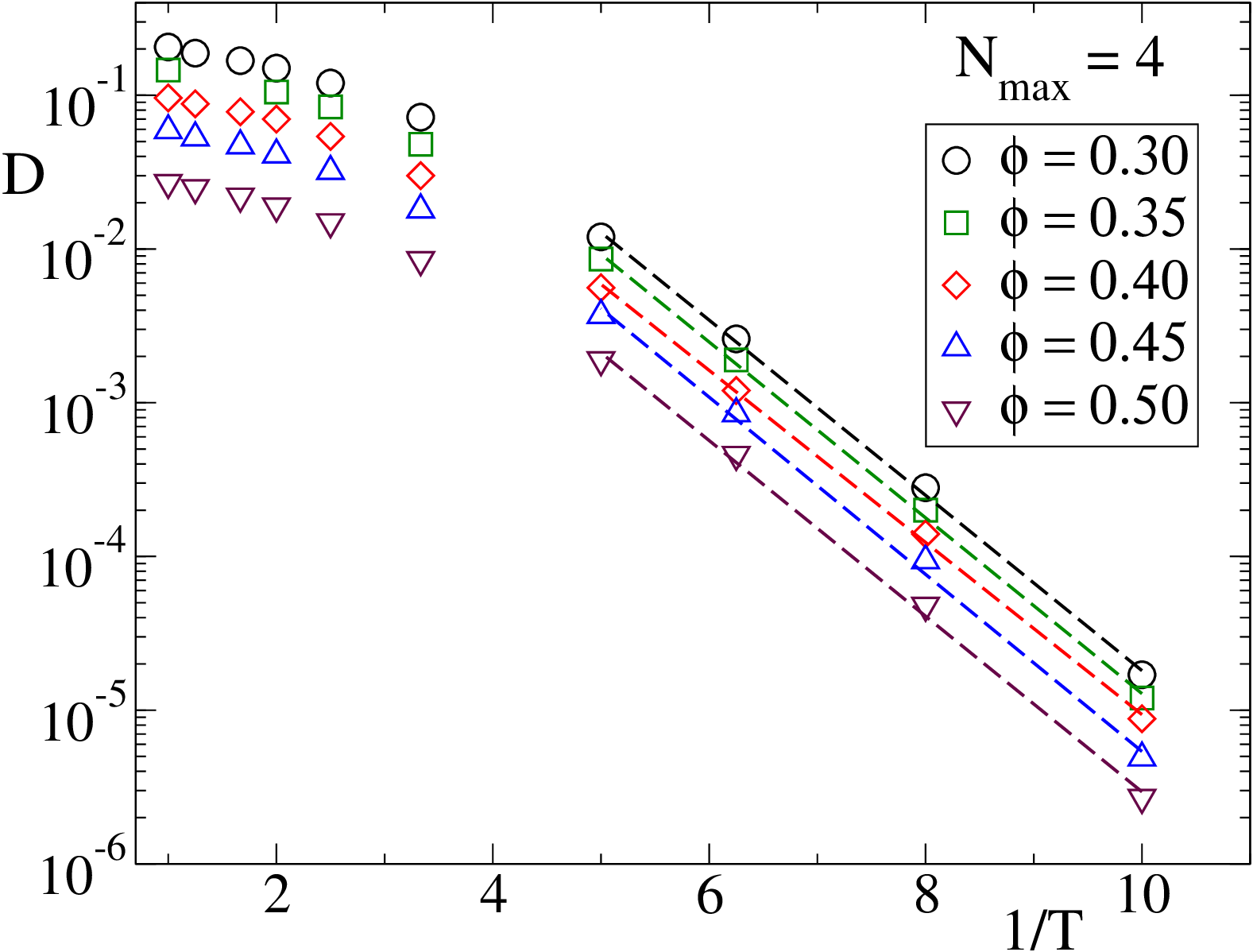}
\newline
\newline
\includegraphics[width=.4\textwidth]{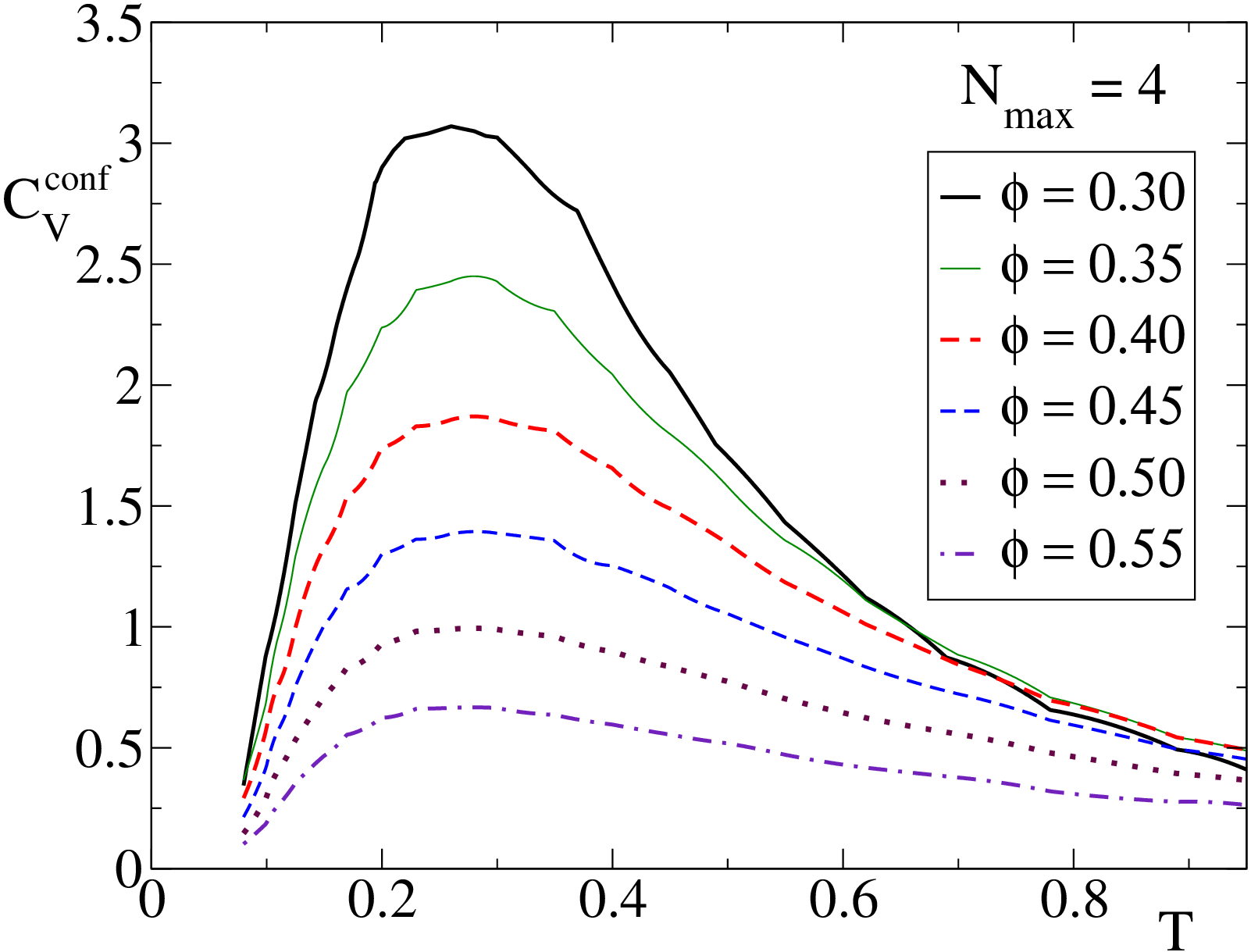}
\newline
\caption{$T$ dependence of the diffusivity (top panel) and the
isochoric configurational specific heat (bottom panel)
for $N_{\rm max}=4$, at several values of $\phi$. Dashed lines in top panel are Arrhenius fits.}
\label{fig:cv}
\end{figure}

The low $T$ Arrhenius dependence of $E$ [Eq. (\ref{eq:arr})] has a practical implication in the
$T$ dependence of the isochoric configurational specific heat $C_{V}^{\rm conf}(T)=(\partial E/\partial T)_{V}$.
Hence, from Eq. (\ref{eq:arr}), at low $T$ we have 
$C_{V}^{\rm conf}(T) = A^{\rm f}_{\infty}\epsilon_{\rm f}k^{-1}_{\rm B}T^{-2}\exp(-\epsilon_{\rm f}/k_{\rm B}T)$,
which has a maximum at $T= \epsilon_{\rm f}/2 \approx u_0/4$. 
Fig. \ref{fig:cvpeak} shows that indeed, numerical data for $C_{V}^{\rm conf}$
display a peak at $T \approx 0.25$.
A peak in $C_{V}^{\rm conf}(T)$ has also been observed  in recent
simulations of atomistic models of two different network-forming liquids: 
silica \cite{voivod01} and BeF$_{2}$ \cite{hemmati}. 

We note that a strong correlation is observed between the $T$ dependence of the diffusivity 
and the $T$ dependence of the potential energy. Fig. \ref{fig:cv} shows, for $N_{\rm max}=4$
and several $\phi$ values, that on cooling, $D$ crosses to an Arrhenius law at $T\approx 0.25$.  This temperature
is the same at which the specific heat shows a maximum. This correlation holds at all the investigated $\phi$ values.

The quality of the low $T$ Arrhenius fits for the diffusivities is worse at high $\phi$.
Indeed, low $T$ data at high $\phi$ show some bending (Fig. \ref{fig:cv}a). This result
suggests that the system will become more fragile with increasing density. This is not surprising,
since the influence of the square well will be weaker at higher packing, and
the system will approach a dense hard sphere liquid, which is a fragile system.

\begin{figure}
\includegraphics[width=.43\textwidth]{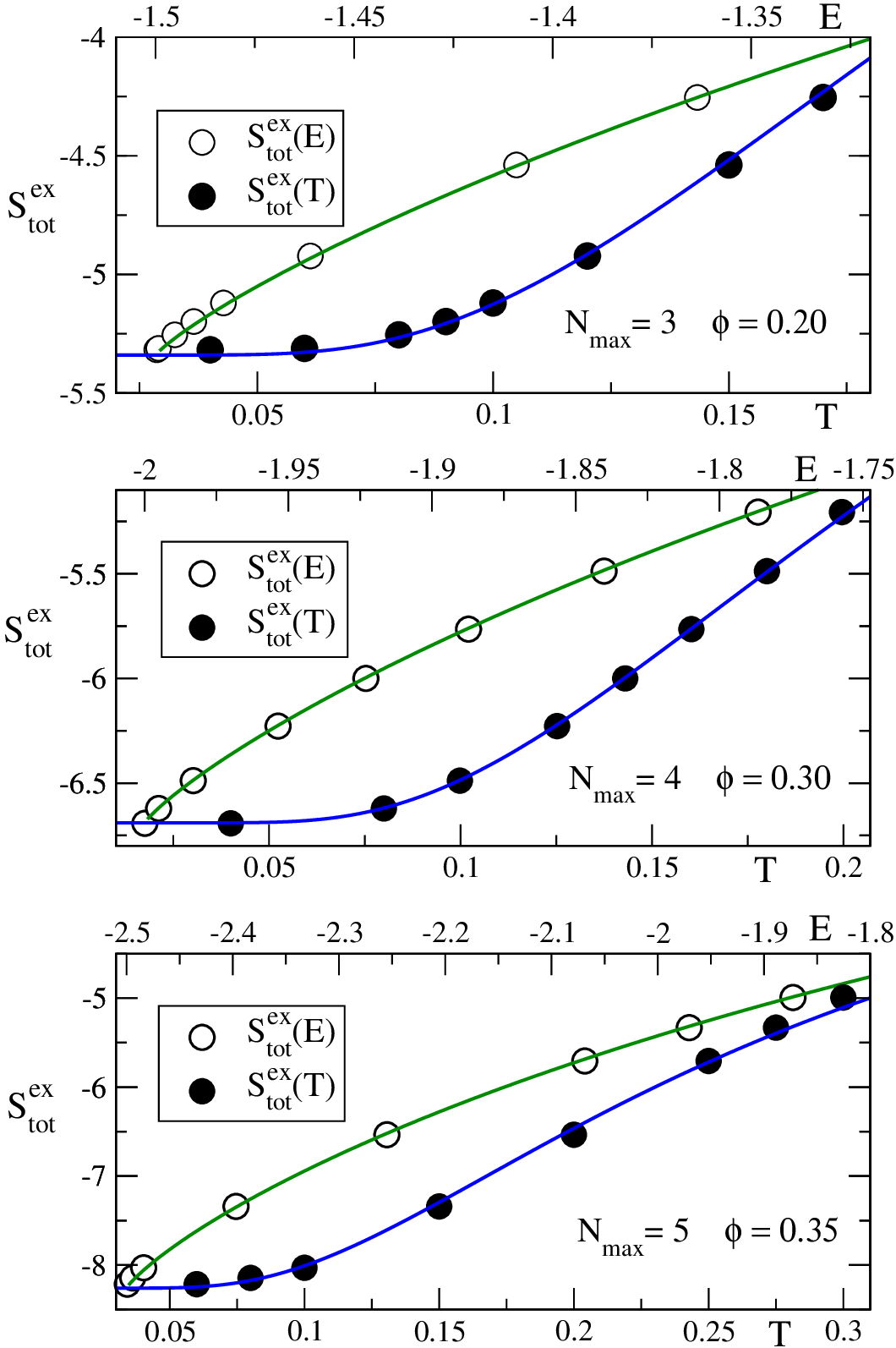}
\caption{$E$  and $T$ dependence of the total excess entropy over the ideal gas
value, $S^{\rm ex}_{\rm tot}$, for the cases  $N_{\rm max} = 3$, $\phi = 0.20$ (top panel),
$N_{\rm max} = 4$, $\phi = 0.30$ (middle panel), and $N_{\rm max} = 5$, $\phi = 0.35$ (bottom panel).
Continuous lines for $S^{\rm ex}_{\rm tot}(E)$ are obtained as the sum of the fit functions for
$S^{\rm ex}_{\rm vib}(E)$ [Eq. (\ref{eq:fitsvibex})] and $S_{\rm conf}(E)$ [Eq. (\ref{eq:sconf})].
Continuous lines for $S^{\rm ex}_{\rm tot}(T)$ are parametrically obtained from the $T$ dependence of $E$.}
\label{fig:stot}
\end{figure}

Next we turn to the evaluation of the statistical properties of the PEL,
and more precisely the total excess entropy and its vibrational and configurational contributions. 
As explained in Section III [Eq. (\ref{eq:stotex})], the total excess entropy $S_{\rm tot}^{\rm ex}$ is evaluated
by isochoric integration from the hard sphere limit at a sufficiently high $T_{\rm ref}$. 
We use a value $T_{\rm ref}=100$. Fig. \ref{fig:stot} shows $S_{\rm tot}^{\rm ex}$ as a function of $T$ and $E$, 
for different values of $\phi$ and $N_{\max}$. Due to the presence of the interaction 
potential (\ref{eq:model1}, \ref{eq:model2}), $S_{\rm tot}^{\rm ex}$ is negative (see also below).
As expected, $S^{\rm ex}_{\rm tot}(T)$ decays to a constant value at low $T$, since 
the system is very close to the fully bonded state and no further structural changes are expected to
occur. A glass transition temperature $T_{\rm g}$ can be operationally defined as 
the $T$ at which relaxation becomes longer than the simulation time.
The fact that the system is so close to its lowest energy state already above this operational $T_{\rm g}$
implies that a very small drop of the specific heat is expected at $T_{\rm g}$, consistently with 
experimental evidence in strong liquids \cite{martinez}.

\begin{figure}
\includegraphics[width=.41\textwidth]{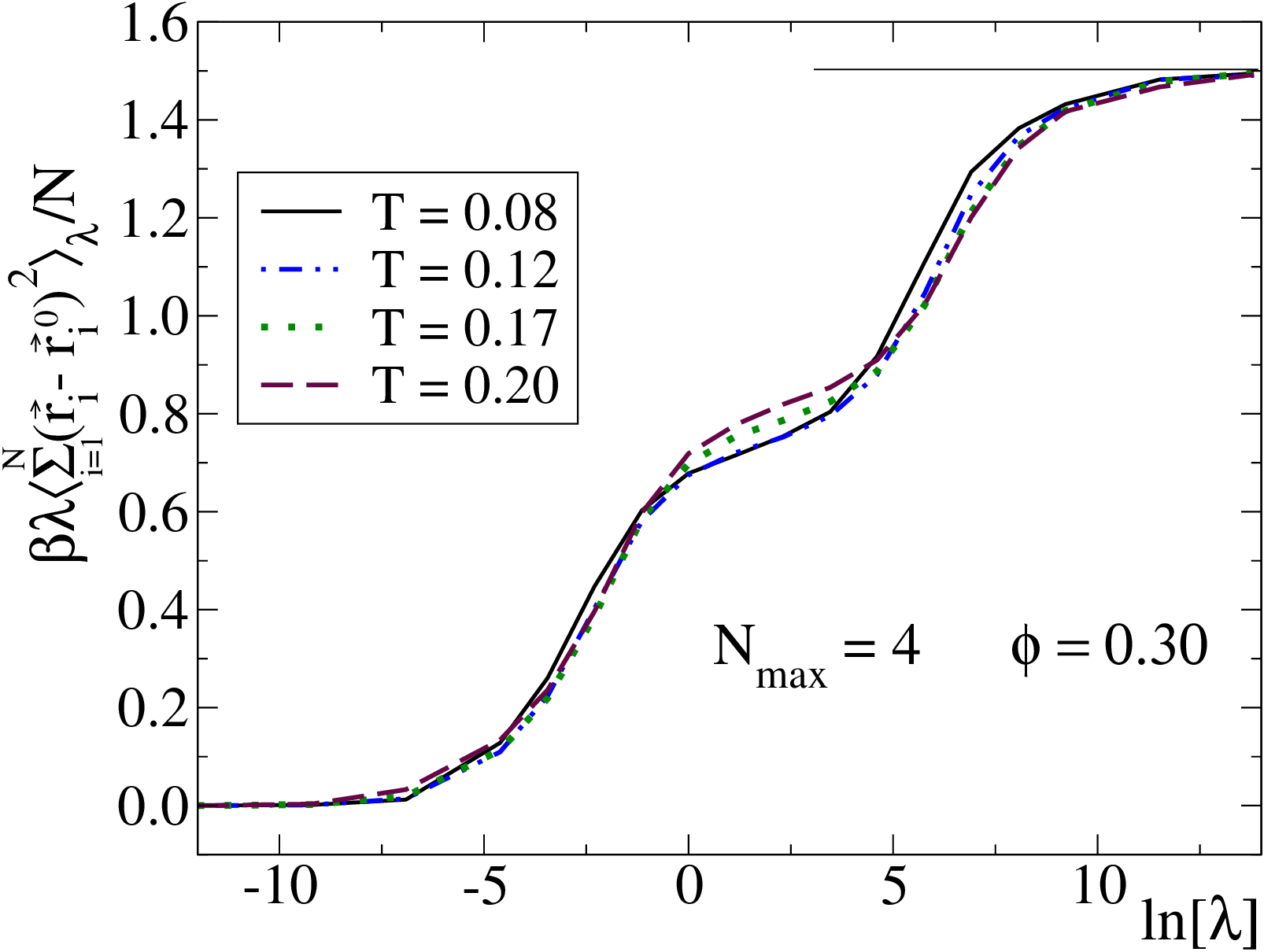}
\caption{$\lambda$ dependence of 
$\beta\lambda  \langle \sum_{i=1}^N ({\bf r}_i - {\bf r}_i^0)^2 \rangle_{\lambda} /N $, 
for $N_{\rm max} = 4$ and $\phi = 0.30$, at different temperatures.
The horizontal line indicates the expected value 3/2 for harmonic behavior.}
\label{fig:lammsd}
\end{figure}

The evaluation of the vibrational contribution  $S_{\rm vib}^{\rm ex}$ 
[Eq. (\ref{eq:svibex})] requires  the calculation of the  integral over the coupling constants $\lambda$.
Fig. \ref{fig:lammsd} shows the calculated $\lambda$ dependence
of $\lambda  \langle \sum_{i=1}^N ({\bf r}_i - {\bf r}_i^0)^2 \rangle_{\lambda} /N $ at 
several $T$ for one specific value of $N_{\rm max}$ and $\phi$. Data for different $N_{\rm max}$ and
$\phi$ display a similar behavior.  At values $\lambda < \lambda_{\rm min} \approx 10^{-6}$ 
contributions to the integral in Eq. (\ref{eq:svibex})  are negligible, and $\lambda_{\rm min}$ is taken as lower-cut
for integration. At large  $\lambda$ values, 
$ \lambda  \langle \sum_{i=1}^N ({\bf r}_i - {\bf r}_i^0)^2 \rangle_{\lambda} /N $  
approaches the theoretical limit $3k_{\rm B}T/2$.
This value is reached at $\lambda_{\rm max}  \gtrsim 10^{6}$.  
Note that $\lambda=10^{6}$ corresponds to an
average displacement per particle of the order of $10^{-4} $ for this $T$ range. Hence the harmonic
perturbation localize the particles in a well much narrower than the square well width $\Delta$,
so that the presence of the unperturbed potential is irrelevant for this upper $\lambda$ value.

\begin{figure}
\includegraphics[width=.43\textwidth]{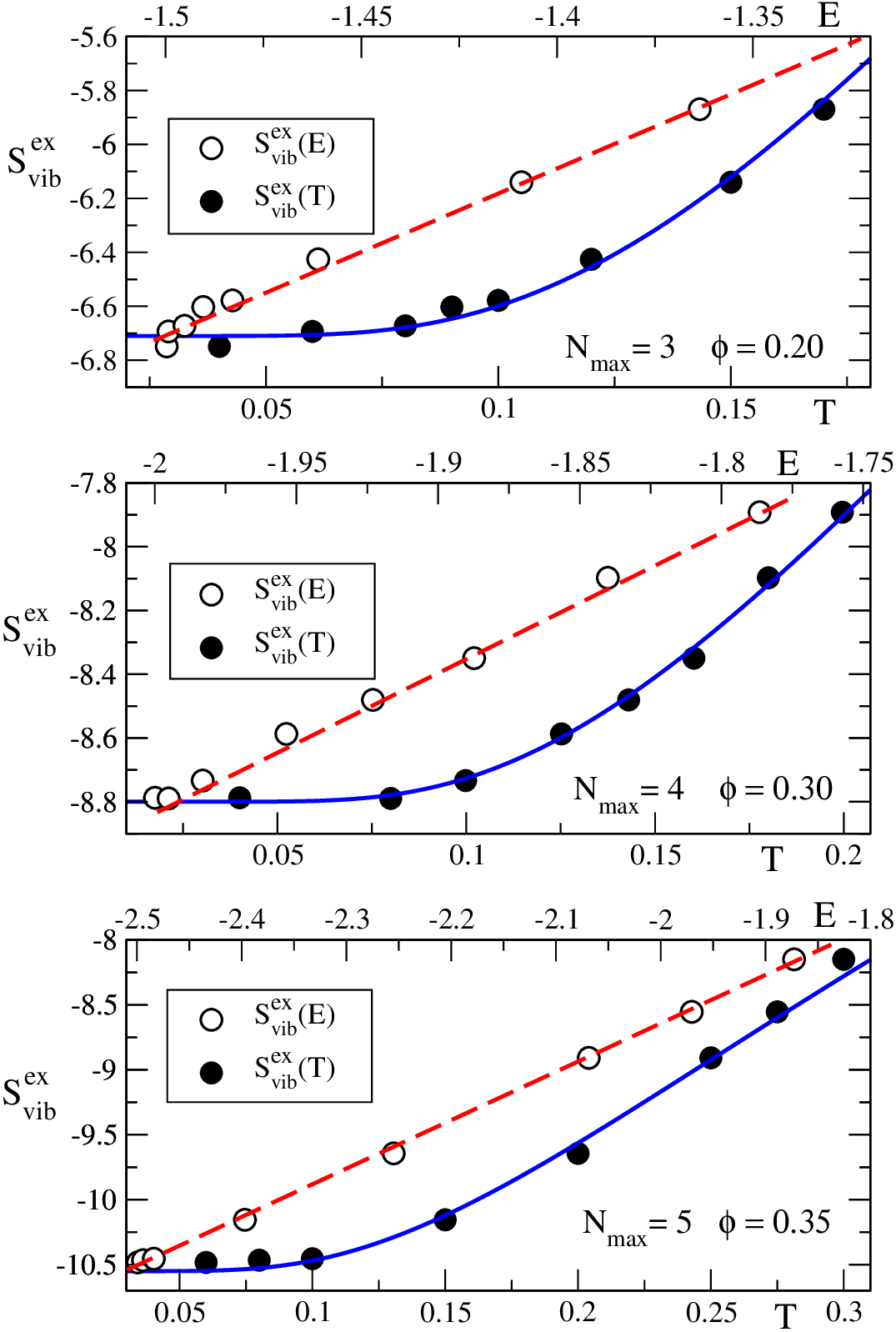}
\caption{As Fig. \ref{fig:stot} for the excess vibrational entropy, $S^{\rm ex}_{\rm vib}$.
Dashed lines for $S^{\rm ex}_{\rm vib}(E)$ are linear fits [Eq. (\ref{eq:fitsvibex})].
Continuous lines for $S^{\rm ex}_{\rm vib}(T)$
are parametrically obtained from the $T$ dependence of $E$ (see text).}
\label{fig:svib}
\end{figure}

Fig. \ref{fig:svib} shows the $T$  and $E$  dependence of $S_{\rm vib}^{\rm ex}$  for the same
values of $\phi$ and $N_{\rm max}$ of Fig. \ref{fig:stot}.  Close to the fully connected state
$S_{\rm vib}^{\rm ex}$ can be well described by a linear dependence on the energy (i.e. on the number of
bonds):
\begin{equation}
S^{\rm ex}_{\rm vib}(E) = S^{\rm ex}_{\rm vib}(E_{\rm fb}) + \gamma_{\rm vib}(E-E_{\rm fb}).
\label{eq:fitsvibex}
\end{equation}
Interestingly, this linear dependence on the basin depth has also been observed in previously investigated
models of supercooled liquids \cite{sastry01,press,mossaotp,buchner1,wales}.

\begin{figure}
\includegraphics[width=.43\textwidth]{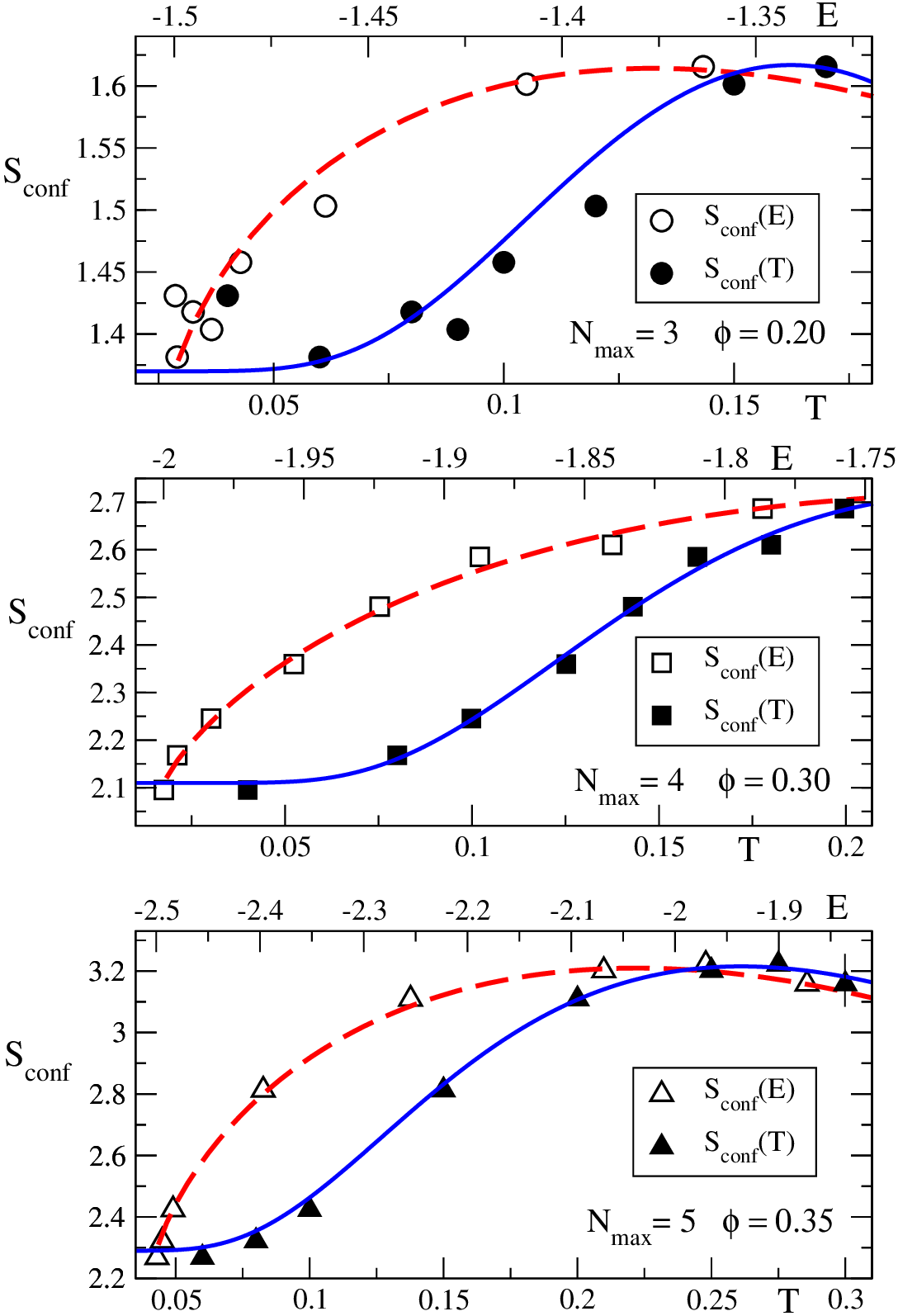}
\caption{As Figs. \ref{fig:stot} and \ref{fig:svib} for the configurational entropy, $S_{\rm conf}$.
Dashed lines for $S_{\rm conf}(E)$ are fits to Eq. (\ref{eq:sconf}).
Continuous lines for $S_{\rm conf}(T)$
are parametrically obtained from the $T$ dependence of $E$ (see text).
A typical error bar is shown in down panel.}
\label{fig:sconf}
\end{figure}

Finally, the configurational entropy $S_{\rm conf}$ is determined as  $S^{\rm ex}_{\rm tot}-S^{\rm ex}_{\rm vib}$.
Fig. \ref{fig:sconf} shows  $S_{\rm conf}$ for the same $\phi$ and $N_{\rm max}$ of Figs. \ref{fig:stot} and \ref{fig:svib}.
Some considerations are in order: as for the total entropy,  $S_{\rm conf}$ approaches a constant value
at low $T$. Interestingly enough, this constant value is significantly different from zero. The fully connected
network is thus characterized by an extensive number of distinct bond configurations $\sim \exp(N S_{\rm conf})$.
These different network configurations arise from different bond topologies, 
i.e., disorder is associated to the presence of closed loops of different number of bonds.  

To derive a functional form for the $E$ dependence of $S_{\rm conf}$, we start from the thermodynamic
relation $(\partial S/\partial E)_V =1/T$, which in the present case can be written as:
\begin{equation}
\frac{\partial(S_{\rm conf}+S_{\rm vib})}{\partial(E-E_{\rm fb})}=\frac{1}{T}.
\end{equation}
At low $T$, from Eq. (\ref{eq:arr}) we obtain 
$1/T= -(1/\epsilon_{\rm f}) \ln [(E-E_{\rm fb})/A^{\rm f}_\infty]$ and,
making use of Eq. (\ref{eq:fitsvibex}) we find:
\begin{equation}
\frac{\partial S_{\rm conf}}{\partial(E-E_{\rm fb})}= 
-\gamma_{\rm vib} -\frac{1}{\epsilon_{\rm f}} \ln \frac{E-E_{\rm fb}}{A^{\rm f}_\infty}, 
\end{equation}
which, after integration, provides the $E$ dependence of the configurational entropy:
\begin{eqnarray}
\nonumber S_{\rm conf}(E)=S_{\rm conf}(E_{\rm fb}) \\ 
-\frac{E-E_{\rm fb}}{\epsilon_{\rm f}}\ln[2(E-E_{\rm fb})] +\gamma_{\rm conf}(E-E_{\rm fb}),
\label{eq:sconf}
\end{eqnarray}
where the constant $\gamma_{\rm conf}$ is given by:
\begin{equation}
\gamma_{\rm conf} =\frac{1}{\epsilon_{\rm f}} -\gamma_{\rm vib} +\frac{\ln(2A^{\rm f}_\infty)}{\epsilon_{\rm f}}.
\label{eq:gconf}
\end{equation}
As mentioned above, a satisfactory description of the low $T$ Arrhenius dependence of $E$
is provided by forcing a value  $u_0/2$ for the activation energy $\epsilon_{\rm f}$. 
Hence, we can make the changes
$\epsilon_{\rm f} \rightarrow u_0/2$ and $A^{\rm f}_\infty \rightarrow A_\infty$
in Eqs. (\ref{eq:sconf}, \ref{eq:gconf}).
These changes allow us to obtain a simple expression of $S_{\rm conf}$ in terms of the 
number of broken bonds. Hence, since $E-E_{\rm fb} = N_{\rm bb}u_0$, we find:
\begin{equation}
S_{\rm conf}(E) = S_{\rm conf}(E_{\rm fb}) -2N_{\rm bb}\ln(2 N_{\rm bb}) +\gamma_{\rm conf}N_{\rm bb}.          
\label{eq:sconfnbb}
\end{equation}     
This expression suggests that the low-$E,T$ dependence of the configurational entropy is controlled by
a combinatorial factor  related to the number of  broken bonds  randomly distributed along the network.
The derivation also shows how intimately the logarithmic dependence of the entropy  is connected
to the Arrhenius dependence of $E$ at low $T$. The final expression for  $S_{\rm conf}(E)$ 
is very different from the quadratic energy dependence  resulting from the Gaussian distribution of IS energies
observed in models of fragile liquids \cite{st,sastry01,starr01,press}.

\begin{figure}
\includegraphics[width=.4\textwidth]{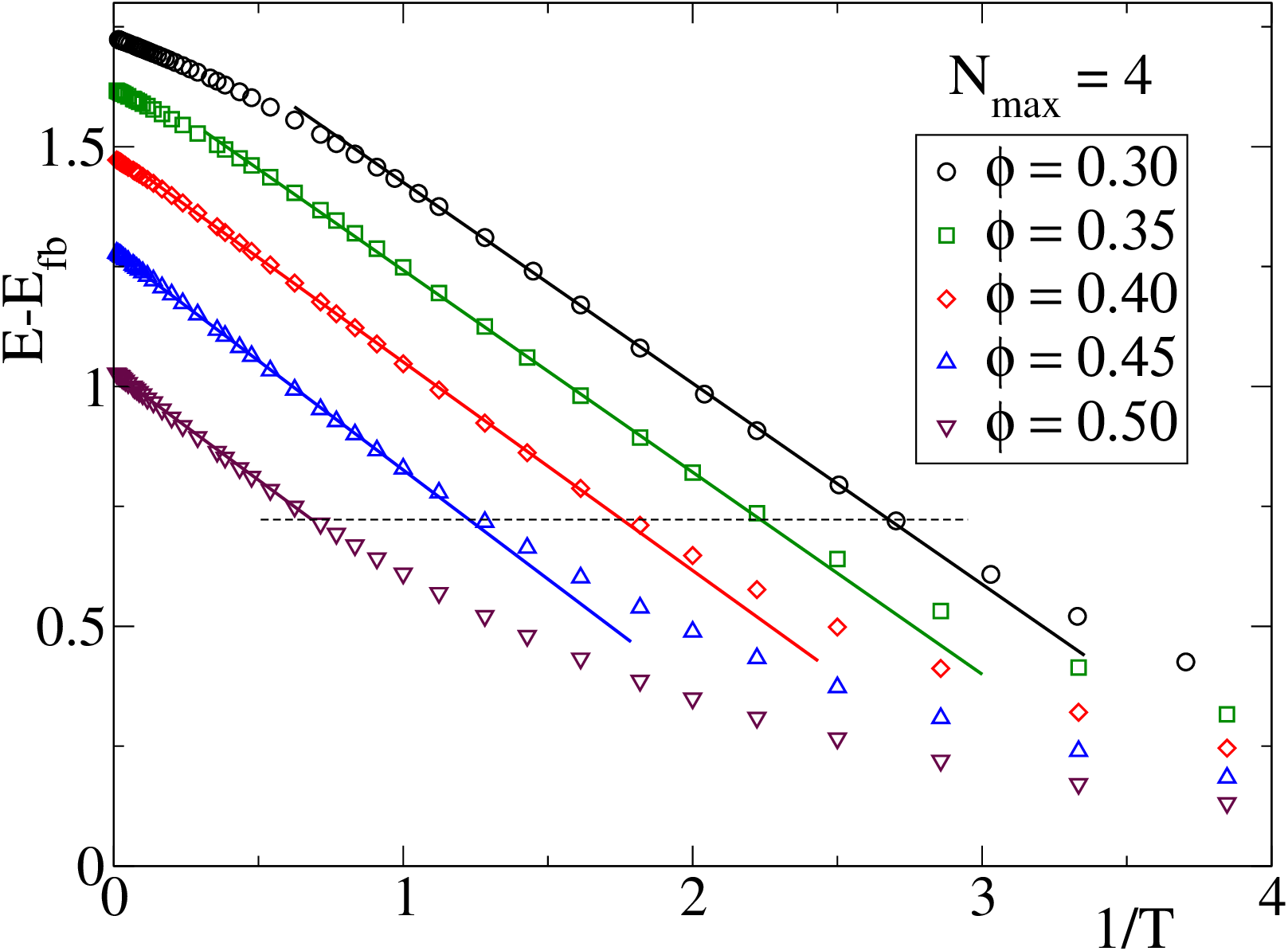}
\newline
\newline
\includegraphics[width=.4\textwidth]{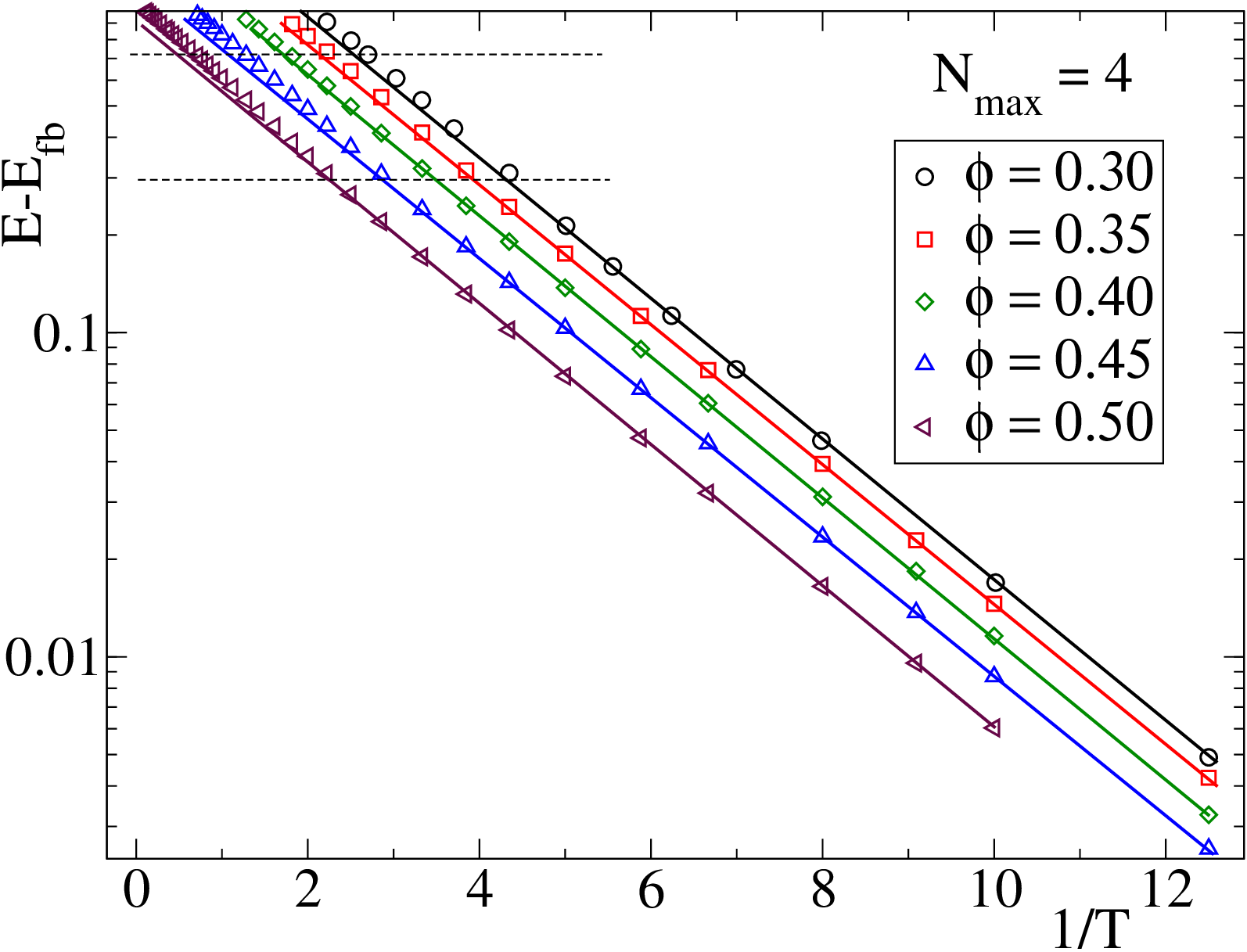}
\newline
\caption{$T$ dependence of $E-E_{\rm fb}$ at different regions of the energy landscape.
Top panel: Full lines correspond to behavior linear in $1/T$. The horizontal dashed line
indicates the departure of such behavior. Bottom panel: Full lines correspond to Arrhenius behavior.
The horizontal dashed lines indicate the limits of Arrhenius and $1/T$ behavior.}
\label{fig:gautoarr}
\end{figure}

To provide a further and analysis-free confirmation of the crossover at low $T$ toward 
combinatorial statistics of the bonding energy states, we show in Fig. \ref{fig:gautoarr}
the $T$ dependence of $E$ for $N_{\rm max}=4$ at several $\phi$ values. 
We show a representation of $E-E_{\rm fb}$, both in linear and logarithmic scale, as a function of $1/T$. We note
that, for all $\phi$, when $E-E_{\rm fb} \approx 0.75$ the $1/T$ law, which is expected to hold in a
Gaussian landscape, breaks down. Similarly, when  $E-E_{\rm fb} \approx 0.3$ the Arrhenius law 
sets in. The fact that both crossovers are, at most, weakly dependent on $\phi$ suggests that they are 
esentially controlled by the bonding statistics and that 
between $E-E_{\rm fb} \approx 0.75$ and $E-E_{\rm fb} \approx 0.3$
a crossover from Gaussian to logarithmic statistics occurs.

\begin{figure}
\includegraphics[width=.45\textwidth]{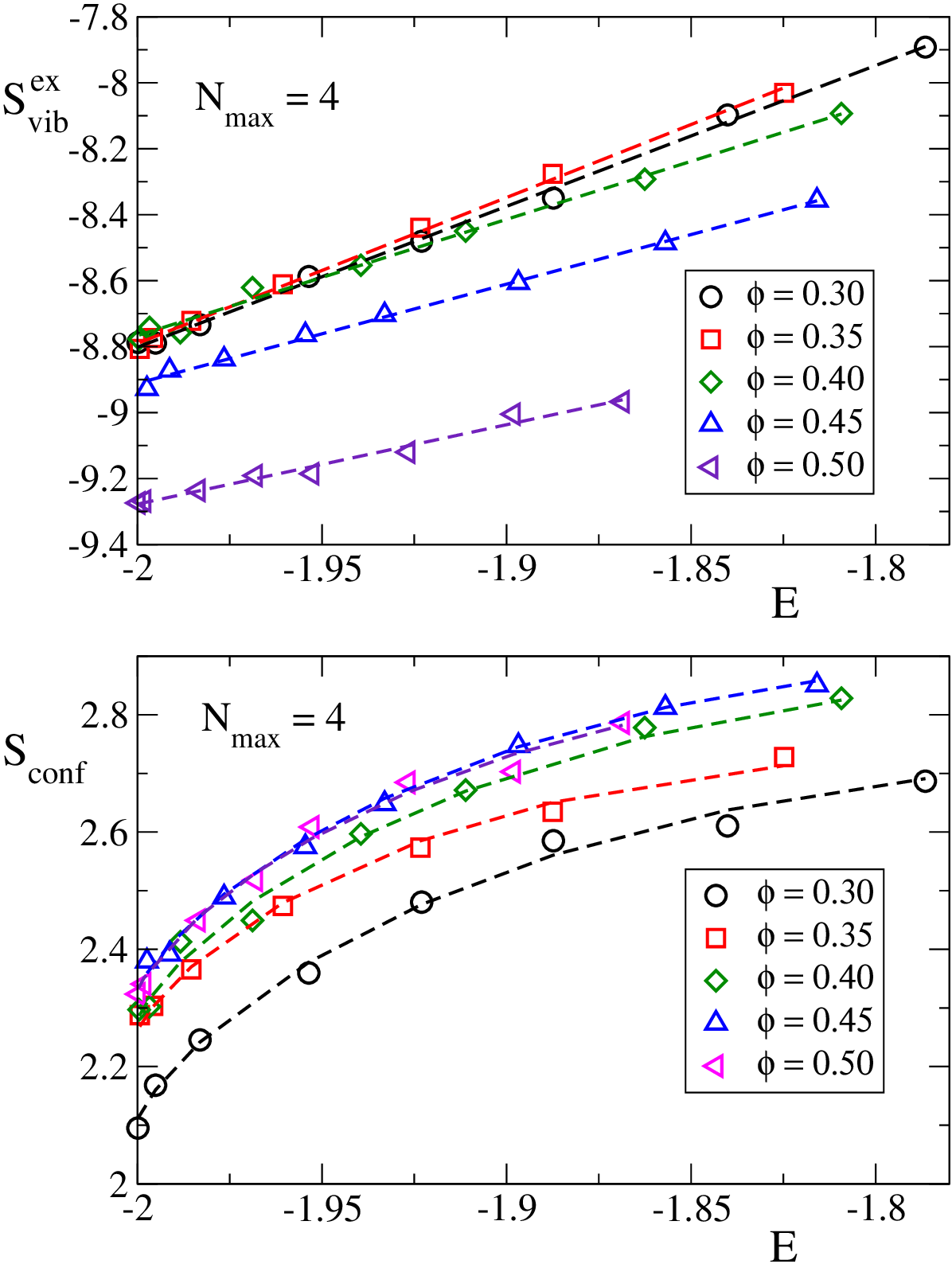}
\newline
\caption{$E$ dependence of $S_{\rm vib}^{\rm ex}$ and $S_{\rm conf}$
for $N_{\rm max}=4$ and different values of $\phi$.
Dashed lines in top and bottom panels are, respectively,
fits to Eqs. (\ref{eq:fitsvibex}) and (\ref{eq:sconf}).}
\label{fig:phiuscsv}
\end{figure}

Fig. \ref{fig:sconf} shows the $E$ and $T$ dependence of $S_{\rm conf}$ for different values of $N_{\rm max}$.
Eq. (\ref{eq:sconf}) provides a good description of the data. No fit parameters are involved
in comparing the numerical estimates of $S_{\rm conf}$ and the predictions of  Eq. (\ref{eq:sconf}) , except for
the constant $S_{\rm conf}(E_{\rm fb})$. Note that $\gamma_{\rm conf}$ is not a fit parameter,
but a function [Eq. (\ref{eq:gconf})] of parameters defining $E(T)$ and $S^{\rm ex}_{\rm vib}(E)$. 
Before discussing the calculated values for $S_{\rm conf}(E_{\rm fb})$, we present in Fig. \ref{fig:phiuscsv}
the $E$ dependence of the configurational and excess vibrational 
entropies for $N_{\rm max}=4$ and different packing fractions $\phi$.
For all $\phi$, the same functional forms of Eqs. (\ref{eq:fitsvibex}) and (\ref{eq:sconf}) are recovered 
respectively for $S^{\rm ex}_{\rm vib}$ and $S_{\rm conf}$.
The two quantities show an opposite trend. Curves for $S_{\rm conf}(E)$ tend to collapse at
high $\phi$, while those for $S^{\rm ex}_{\rm vib}(E)$ tend to collapse at low $\phi$.
Interestingly, the configurational entropy just shifts with varying $\phi$.

Table \ref{tab:entro} summarizes the results of the fits of the vibrational and configurational entropies
to Eqs. (\ref{eq:fitsvibex}) and (\ref{eq:sconf}) for the studied range of control parameters.
Data shown in the table help discussing the $N_{\rm max}$ and
$\phi$ dependence of the entropy of the fully bonded state. A trend in the direction of increasing
$S_{\rm conf}(E_{\rm fb})$ on increasing $N_{\rm max}$ is observed very clearly in the
comparison between $N_{\rm max}=3$ and $N_{\rm max}=4$. Much weaker is the trend 
between $N_{\rm max}=4$ and $N_{\rm max}=5$. A similar weak trend is observed for
$N_{\rm max}=4$ on increasing $\phi$. 
The weak increase of  $S_{\rm conf}(E_{\rm fb})$ with increasing $\phi$ suggests that when 
the system is compressed, neighboring particles progressively enter in the interaction range
of a given one, yielding a major variety of local configurations of the bonding pattern,
and consequently, more topologically distinct fully bonded networks.
The trend of the excess vibrational entropy suggests that the  
available free volume for a given bonding pattern decreases with increasing $\phi$.

\begin{table}
\begin{center}
\begin{tabular}{lcccccccc}
&$(N_{\rm max},\phi)$  &$S_{\rm conf}(E_{\rm fb})$  &$\gamma_{\rm conf}$ 
&$S^{\rm ex}_{\rm vib}(E_{\rm fb})$  &$\gamma_{\rm vib}$  \\
\hline
\\
& (3, 0.20)     & 1.37   & -0.79 & -6.71   & 6.16    \\
\vspace{0.2mm}
& (3, 0.30)     & 1.55   &  -0.336  & -6.60   & 4.83    \\
\vspace{0.2mm}
& (3, 0.35)     & 1.63   &  -1.02  & -6.70   & 5.05    \\
\vspace{0.2mm}

& (4, 0.30)     & 2.11   & 1.02 & -8.80   & 4.27    \\
\vspace{0.2mm}
& (4, 0.35)     & 2.26   & 0.49 & -8.79   & 4.43    \\
\vspace{0.2mm}
& (4, 0.40)     & 2.28   & 0.93 & -8.77   & 3.52    \\
\vspace{0.2mm}
& (4, 0.45)     & 2.33   & 0.87 & -8.91   & 3.01    \\
\vspace{0.2mm}
& (4, 0.50)     & 2.33   & 0.77  & -9.27   & 2.41    \\
\vspace{0.2mm}
& (5, 0.35)     & 2.28   & 1.84  & -10.55   & 3.83 \\    
\\
\hline
\end{tabular}
\end{center}
\caption{Parameters defining the configurational [Eq.(\ref{eq:sconf})]
and excess vibrational [Eq.(\ref{eq:fitsvibex})] entropy for the studied values
of $N_{\rm max}$ and $\phi$.}
\label{tab:entro}
\end{table}

\begin{figure}
\includegraphics[width=.36\textwidth]{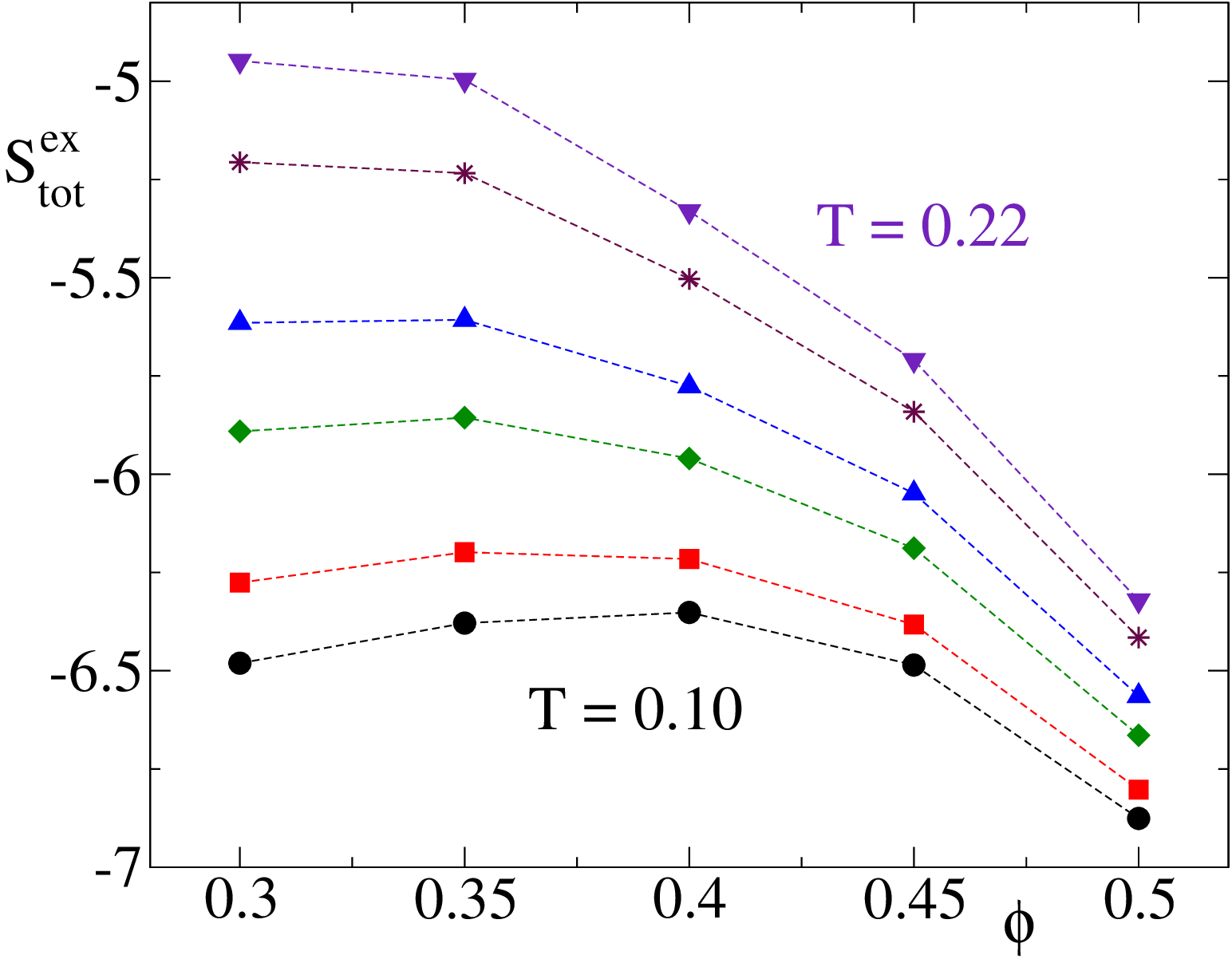}
\newline
\newline
\includegraphics[width=.36\textwidth]{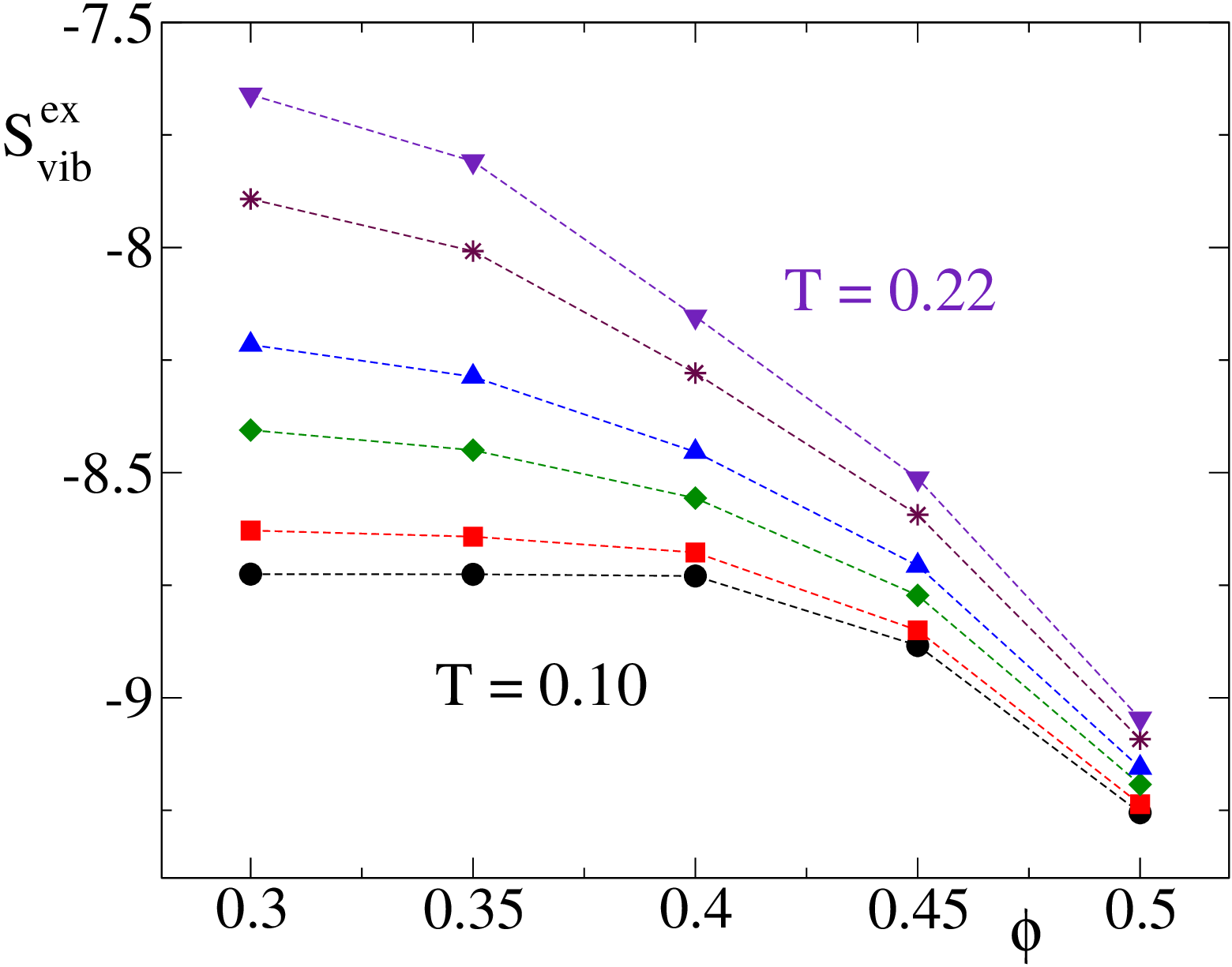}
\newline
\newline
\includegraphics[width=.36\textwidth]{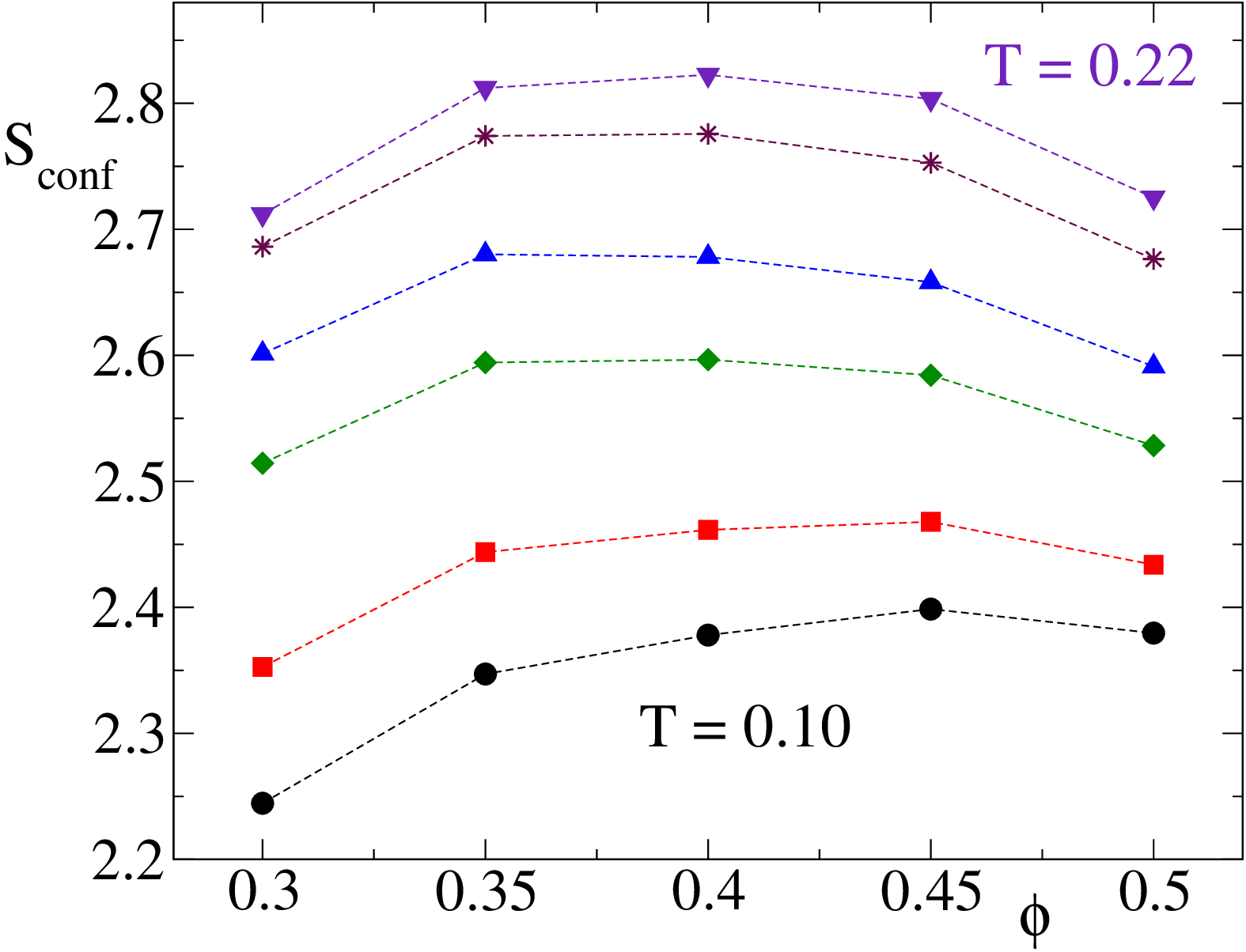}
\newline
\newline
\includegraphics[width=.36\textwidth]{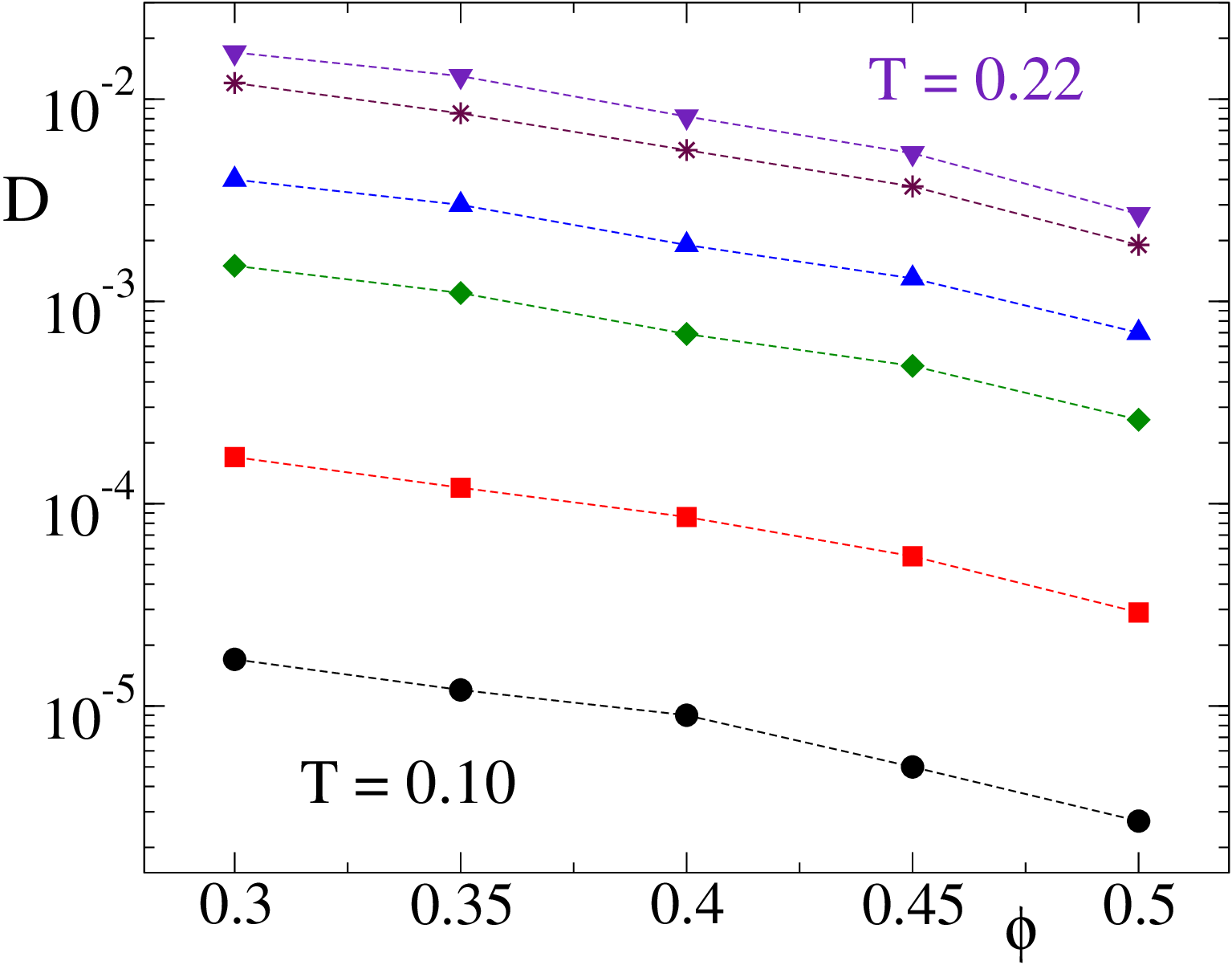}
\newline
\caption{$\phi$ dependence of $S^{\rm ex}_{\rm tot}$, $S^{\rm ex}_{\rm vib}$, $S_{\rm conf}$, and $D$,
for $N_{\rm max}=4$ along isothermal curves. In all figures, from top to bottom, 
the isothermals are $T=$ 0.22, 0.20, 0.17, 0.15, 0.12 and 0.10. Dashed lines are guides for the eyes.}
\label{fig:isotherm}
\end{figure}

A summary of the landscape analysis for all studied $\phi$ is shown in Fig. \ref{fig:isotherm} for $N_{\rm max}=4$.
The figure shows the $\phi$ dependence of $S_{\rm tot}^{\rm ex}$, $S_{\rm vib}^{\rm ex}$ and $S_{\rm conf}$   
for several different low $T$ isotherms, all in the Arrhenius region of energies. 
It clearly emerges that the significant reduction of $S_{\rm tot}^{\rm ex}$ on
increasing $\phi$ arises essentially from the vibrational component. We also note 
that the HS relative contribution to $S_{\rm tot}^{\rm ex}$ increases on increasing $\phi$. Hence, according
to Eq. (\ref{eq:carnstar}), $S^{\rm ex}_{\rm HS}(\phi=0.30) = -1.90$, while  
$S^{\rm ex}_{\rm HS}(\phi=0.50) = -5.0$. By comparing with data in Fig. \ref{fig:isotherm}, it is clear
that $S_{\rm tot}^{\rm ex}$ is dominated by the square-well and hard-sphere contributions at respectively
low and high packing fraction.

It is also interesting to observe that, within the
precision of the data, $S_{\rm conf}$  shows a weak maximum, shifting to higher $\phi$ on cooling.  
To test wether  the presence of a maximum of $S_{\rm conf}$  has some effect on the dynamics
we also show in Fig. \ref{fig:isotherm} the behavior of the diffusivity along the same isotherms. We note that
$D$ is monotonic in $\phi$ and hence that the maximum in the configurational entropy does not provoke a maximum in
the diffusivity. We also note that an isochoric plot (not shown) of $\log D$ vs. $[TS_{\rm conf}(T)]^{-1}$ 
provides a satisfactory linearization of the data, as suggested by the Adam-Gibbs theory \cite{adamgibbs}. 
This is not inconsistent with the observed Arrhenius dependence of $D$, 
since the $T$ dependence of $S_{\rm conf}(T)$ is only at most 20\% of   $S_{\rm conf}(E_{\rm fb})$.

\begin{center}
{\bf V. CONCLUSIONS}
\end{center}

This article reports an explicit numerical calculation of the potential energy landscape for
a simple model of strong liquids. It shows that it is possible to calculate with arbitrary precision
the statistical properties of the landscape relevant to the behavior of the system at low $T$,
when all particles are connected by bonds. The model can be seen as a zero-th order model
for network-forming liquids, capturing the limited valency of the interaction and the open structure
of the liquid. By construction, it misses all  geometric correlations between different bonds 
which are present in  network-forming materials. The  simplicity of the model has several
advantages, some of which of fundamental importance for an exact evaluation of the landscape
properties.  Hence, since angular constraints between bonds are missing, it is possible to equilibrate the
system to very low $T$, reaching configurations which are essentially fully bonded. At the lowest studied
$T$, less than 2\% of the bonds are broken in average. Differently from other studied models with 
fixed bonding site geometries \cite{kolafa,vega,demichele}, the absence of geometric constraints makes it
possible to reach almost fully bonded states in a wide range of densities. Moreover,
the use of a square well interaction as bonding potential has the advantage that the energy of the
fully bonded state (the lowest possible energy) is known. 

The present model neglects completely interactions between particles which are not
nearest neighbors. The energy of a particle is indeed fully controlled by the bonds with the
nearby particles. This element of the model favors a sharp definition of the energy and a clear-cut
definition of basins. On the other hand, in network-forming liquids, bonding is often 
of electrostatic origin and interactions are not limited to the first shell of neighbors. This produces
a much wider variety of local environments and, as a consequence, a spreading of the energy levels.
These residual interactions, if smaller than the bonding interactions, will only contribute to 
spread the distribution of energy states without changing the landscape features 
(down to $T$ of the order of the energy spreading).    

The use of square-well interactions has a major advantage in relation to the possibility of  precisely calculating
landscape properties, since the energy of the system becomes a measure of the number of bonds.
A basin in configuration space can be identified as a bonding pattern, and a transition between
different basins becomes associated to bond forming and breaking. Under these conditions, the basin 
boundaries are properly defined. We have shown that the method of the Perturbed Hamiltonian can be
extended to the present model, providing a formally exact method to evaluate the vibrational
component of the free energy.  This is a relevant achievement, since the evaluation of the vibrational entropy
is the only weak point in all estimates of landscape properties
in models with continuous potentials \cite{jstat,angelanientro}. 
Indeed, when the potential is continuous, the constraint of exploring a fixed basin
can not be implemented unambiguously in the Frenkel-Ladd method due to the difficulty
of detecting crossing between different basins.
Such a difficulty is not present in the square-well potential, 
since crossing of a basin  is detected by a finite energy change.

The two relevant features observed in this study are: (i) A residual value of the configurational entropy for
$T \rightarrow 0$, associated to the exponentially large number of distinct fully connected bonding patterns.
(ii) A logarithmic dependence of the number of bonding patterns on the number of broken bonds [Eq. (\ref{eq:sconfnbb})].
These two features are common to all investigated values of $N_{\rm max}$  and to all studied $\phi$. 
A consequence of the logarithmic landscape statistics is the absence of a finite temperature
at with the lowest energy state is reached \cite{loglands}.
Indeed from Eq. (\ref{eq:sconf}) it is found that $\partial S/\partial E |_{E=E_{\rm fb}}=\infty$, 
and hence $E_{\rm fb}$ is reached only at $T=0$.

According to the picture emerging from this study, strong liquid behavior
is connected to the existence of an energy scale, provided by the bond energy, 
which is discrete and dominant as compared to the
energetic contributions coming from non-bonded next-nearest neighbor interactions \cite{angellbond}. 
It is also intimately connected to the existence  of a significantly degenerate lowest energy state, 
favoring the formation of highly bonded states which can still entropically rearrange 
to form different bonding patterns with the same energy.  
Of course, the specific value of $S_{\rm conf}(E_{\rm fb})$ in the $N_{\rm max}$ model provides
an upper bound to the value expected in network-forming liquids \cite{sceats,rivier}, 
since the absence of angular constraints 
significantly increases the number of geometric arrangements of the particles compatible with a fully bonded state.
Recent estimates in glassy water \cite{speedywatexp}, which forms a tetrahedral disordered network, 
suggest a residual value of the configurational entropy of the order of $\approx 0.3k_{\rm B}$ .
Hence the localization of the bonding sites at specific locations, and the associated geometrical correlations,
do produce a significant reduction of $S_{\rm conf}(E_{\rm fb})$. 

Results reported in this article suggest that strong and fragile liquids are characterized
by significant differences in their potential energy landscape properties.   
A non-degenerate disordered lowest energy state and Gaussian statistics characterize fragile liquids, 
while a degenerate disordered lowest energy state and logarithmic statistics are associated with strong liquids.
These results rationalize previous landscape analysis of realistic models of
network-forming liquids \cite{voivod01}, and the recent observation by Heuer
and coworkers that the breakdown of Gaussian landscape statistics is associated with
the formation of a connected network \cite{newheuer}. 
While in atomistic models the lowest energy state is not known, and the very long equilibration times
prevent an unambiguous determination of its degeneracy, both these quantities are accessible in the
present simple model.

A last remark concerns the limit of the $N_{\rm max}$ value for which the fully connected state can be reached.
The possibility of approaching the fully bonded state is limited by the possibility
of avoiding the $(T-\phi)$ region where liquid-gas
phase separation is present.  It has recently been observed \cite{gel,gellong,gelgenova}
that the region of unstable states expands on increasing
$N_{\rm max}$ and essentially covers, at low $T$, the entire accessible $\phi$ range
when $N_{\rm max} \ge 6$. 
For these large $N_{\rm max}$ values, slowing down of the dynamics is observed only
at very large $\phi$ and it is essentially controlled by
packing considerations, not by bonding. In this respect, bond-controlled dynamics
are observable only when the valence of the interparticle interaction is limited. 
 \\
\begin{center}
{\bf ACKNOWLEDGEMENTS}
\end{center}

We thank  K. Binder,  A. Heuer, C. De Michele, P. H. Poole,
A. M. Puertas, S. Sastry,  R. Schilling, N. Wagner and  F. Zamponi for useful discussions and critical
readings of the manuscript. MIUR-COFIN and MIUR-FIRB are acknowledged  for financial support.
Additional support is also acknowledged from NSF (S. V. B.) and NSERC-Canada (I. S.-V.). 
%
%

%

%


\begin{thebibliography}{99}
\bibitem{anharmonic} This approximation for the anharmonic contribution
works well in the case of soft potentials as Lennard-Jones 12-6 \cite{mossaotp},
SPC/E water \cite{scala}, or BKS silica \cite{voivod01}. However, it breaks down in the
case of short-range steep potentials \cite{angelanientro} as that of the model investigated in this work.
\end{thebibliography}
\end{document}